\documentclass[prd,floatfix, nofootinbib, preprint]{revtex4-1}		%
\usepackage{amssymb}
\usepackage{amsmath}
\usepackage{bm, natbib}
\usepackage{braket}
\usepackage[mathscr]{euscript}
\usepackage{float}
\usepackage[utf8x]{inputenc}
\usepackage{graphicx}
\usepackage{subfigure}
\usepackage{xcolor}

\begin{document}
\title{Favoured $B_{c}$ Decay modes to search for a Majorana neutrino}
\author{Sanjoy Mandal\footnote{smandal@imsc.res.in}, Nita Sinha\footnote{nita@imsc.res.in}}
\affiliation{The Institute of Mathematical Sciences, C.I.T Campus, Tharamani, Chennai 600 113, India.}

\begin{abstract}
Recently, the LHCb collaboration reported the observation of the decay mode $B_{c}^{-}\rightarrow\overline{B}_{s}^{0}\pi^{-}$ with
the largest exclusive branching fraction amongst the known decay modes of all the $B$ mesons. Here we propose a search for a few lepton-number violating ($\Delta L=2$) decay modes of $B_c$ which can only be induced by Majorana neutrinos. Distinguishing between Dirac and Majorana nature of neutrinos is an outstanding problem and hence, all possible searches for Majorana neutrinos need to be carried out. Since the lepton number violating modes are expected to be rare, when using meson decay modes for these searches one expects CKM favoured modes to be the preferred ones; $B_{c}\rightarrow B_{s}$ is one such transition. With a resonance enhancement of the Majorana neutrino mediating the $B_{c}^{-}\rightarrow\overline{B}_{s}^{0}\ell_{1}^{-}\ell_{2}^{-}\pi^{+}$ modes one can hope to observe these rare modes, or, even their non-observation can be used to obtain tight constraints on the mixing angles of the heavy Majorana singlet with the light flavour neutrinos from upper limits of the branching fractions. Using 
these modes we obtain exclusion curves for the mixing angles which are tighter or compatible with results from earlier studies. However, we find that the relatively suppressed mode $B_{c}^{-}\rightarrow J/\psi\ell_{1}^{-}\ell_{2}^{-}\pi^{+}$ can provide even tighter constraints on $\mid V_{e N}\mid^2$, $\mid V_{\mu N}\mid^2$, $\mid V_{e N} V_{\mu N}\mid$, and in a larger range of the heavy neutrino mass. Further, exclusion regions for $\mid V_{e N} V_{\tau N}\mid$, $\mid V_{\mu N} V_{\tau N}\mid$ can also be obtained for masses larger than those accessible in tau decays. Upper limits on $\mathscr{B}\left(B_{c}^{-}\rightarrow\pi^{+}\ell_{1}^{-}\ell_{2}^{-}\right)$ can also result in stringent exclusion curves for all the mixing elements, including that for $\mid V_{\tau N}\mid^2$ in a mass range where it is unconstrained thus far.
\end{abstract}

\maketitle

\section{Introduction}
The discovery of neutrino oscillations~\cite{neutrino oscillation 1,
neutrino oscillation 2,neutrino oscillation 3,neutrino oscillation 4,neutrino oscillation 5,neutrino
oscillation 6,neutrino oscillation 7,neutrino oscillation 8,neutrino oscillation 9,neutrino oscillation 10,
neutrino oscillation 11} confirming the existence of at
least two non-vanishing neutrino mass-squared differences necessitates physics beyond the Standard Model (SM). In principle, neutrino mass could be simply generated by
addition of right-handed (RH) neutrinos through the Higgs mechanism, but to get neutrino masses less than
$1$ eV, the neutrino Yukawa coupling has to be extremely small $\sim \mathcal{O}\left(10^{-12}\right)$.
Hence alternate mechanisms for neutrino mass have been proposed. Among these the seesaw mechanism~\cite{seesaw1,seesaw2,seesaw3,seesaw5,Valle 1,Valle 2} provides a natural explanation
of the smallness of neutrino mass. The simplest realization of the seesaw, the so-called type-I seesaw, requires the existence of a set of heavy electroweak singlet
(sterile) lepton number violating (LNV) Majorana fermions, N. A typical scale for the Majorana mass $m_N$ in grand unified theories (GUTs)~\cite{seesaw2} is of the
order of the GUT scale, but in general, in various other scenarios, sterile neutrinos can lie in a wide range of masses. In particular, in low energy seesaw
models~\cite{low energy seesaw1,low energy seesaw2} sterile neutrinos may have mass between $\sim 100$ MeV to few GeV. Sterile neutrinos have also  been invoked
to explain the LSND~\cite{lsnd,lsnd1}, Miniboone~\cite{miniboone,miniboone1,miniboone2} and reactor~\cite{chooz,dayabay,reno} anomalies. A viable dark matter candidate is a KeV sterile
neutrino~\cite{kev sterile1,kev sterile2,kev sterile3,kev sterile4,kev sterile5,kevsterileWP,Queiroz}. Other astrophysical observations including supernovae permit sterile neutrinos mixed with active 
ones. While cosmological/astrophysical constraints on sterile neutrinos are strong, they are model dependent and hence laboratory searches and constraints on sterile
neutrinos, particularly Majorana sterile neutrinos are rather important. Sterile neutrinos have been searched for in the laboratory through peak searches in leptonic
decays of pions and kaons~\cite{peak searches}. The lepton spectrum would show a monochromatic line at a lower energy in presence of a heavy neutrino.
These have provided tight constraints on the mixing angle of the sterile neutrino with the active ones. Heavy neutrinos have also been looked for through searches of
their visible decay products. Searches for sterile neutrinos including majorana sterile neutrinos need to be performed at all possible scales, as their discovery may
provide hints of the new physics responsible for neutrino mass generation. 

One of the promising processes to explore Majorana neutrinos is through neutrinoless double beta decay which may be experimentally feasible due to the large samples
of the decaying nuclei, however, on the theoretical side this involves large uncertainties coming from the nuclear matrix elements making it harder to extract
information on neutrino properties. The rare LNV meson and tau decays can be more accurately evaluated~\cite{tao han,kim2,GammaN calculation} and although
their decay rates may be extremely small, they may be accessible with current and future high luminosity machines. In the last decade or so, many experimental
collaborations, CLEO ~\cite{cleo collaboration1,cleo collaboration2,cleo collaboration3}, FOCUS~\cite{focus}, BaBar~\cite{babar}, BELLE and more recently LHCb~\cite{lhcb1},
have searched for such LNV processes. On the theoretical and phenomenological side as well, considerable effort has been made in proposing possible modes that could probe SM singlet Majorana neutrinos in various mass ranges and
constrain their mixing parameters. This includes proposals to search for heavier neutrinos at accelerator and collider experiments such as,
$e^+e^-$~\cite{Dittmar:1989yg,delAguila:2006bda,Buchmuller:1991tu,Almeida:2000yx,Neutrino and collider physics,Prospects of Heavy Neutrino Searches at Future Lepton Colliders},
$e\gamma$~\cite{delAguila:2006bda,Bray:2005wv},
$pp$ and $p\bar{p}$~\cite{delAguila:2006bda,ppcollider2,ppcollider1,Almeida:2000pz,ppbarcollider1,Neutrino and collider physics,Prospects of Heavy Neutrino Searches at Future Lepton Colliders,New Production Mechanism for Heavy Neutrinos at the LHC,
Direct bounds on electroweak scale pseudo-Dirac neutrinos from 8 TeV LHC data},
$e^-e^-$~\cite{Rizzo:1982kn,Prospects of Heavy Neutrino Searches at Future Lepton Colliders}, as well as in
top quark and W-boson rare decays~\cite{deshpande, fourbody decay}.      

While various $B$, $B_s$ and $B_c$ meson decay modes have already been suggested, here we propose a few additional $B_c$ decay modes that may perhaps be preferable for Majorana neutrino searches. 
The $B_{c}$ mesons are unique, in being the only states consisting of two heavy quarks of different flavours ($b\overline{c}$ for
$B_{c}^{-}$). The weak decay of the b quark will be Cabibbo suppressed, for both $b\rightarrow c$, ($\lambda^{2}$ suppressed)
and $b\rightarrow u$ ($\lambda^{3}$ suppressed) transitions. However, for the $c\rightarrow s$ decay, it will be a Cabibbo favoured transition.
Hence, the mode $B_{c}^{-}\rightarrow\overline{B}_{s}^{0}\ell_{1}^{-}\ell_{2}^{-}\pi^{+}$ is expected to have a larger branching fraction than the other rare lepton number violating decay modes of bottom mesons considered so far.
Further, for a heavy neutrino in the mass range $\sim(0.1-0.9)\,\text{GeV}$, it is kinematically possible for it to be produced as an intermediate on mass shell state, resulting in an additional resonance enhancement of the transition rate. 

The $B_{c}^{-}\rightarrow\overline{B}_{s}^{0}\pi^-$ mode has already been observed by LHCb~\cite{lhcb}. $B_c$ decays to other hadronic modes have also been observed
by ATLAS~\cite{atlas} and CMS~\cite{cms}, hence in addition to LHCb, ATLAS and CMS may also be able to perform the search for Majorana neutrinos via this $B_c$ decay mode.
In the proton-proton collisions at the Large Hadron Collider, $B_c$ mesons are expected to be mainly produced through the gluon-gluon fusion process
$gg\rightarrow B_c^-+\overline{b}+c$~\cite{BcProd}. Hence, the production cross-section would be expected to increase in the $13/14$ TeV run substantially. This, along with the
luminosity of the order of few $\text{fb}^{-1}$ in Run II, leads one to believe that searches for this rare LNV $B_c$ decay modes may be feasible.

In the next section, we give the formalism for the extension of the SM to include right handed singlets. In Sec.~\ref{III}, the four-body decay rate for $B_{c}^{-}\rightarrow\overline{B}_{s}^{0}\ell_{1}^{-}\ell_{2}^{-}\pi^{+}$ mode is evaluated and the expected upper limits on branching ratios for these modes are used to obtain bounds on the mixings of the heavy neutrino with the light flavoured ones. In Sec.~\ref{IV}, the modes $B_{c}^{-}\rightarrow J/\psi\ell_{1}^{-}\ell_{2}^{-}\pi^{+}$ and $B_{c}^{-}\rightarrow \pi^{+}\ell_{1}^{-}\ell_{2}^{-}$ are discussed. We find that although these modes are not Cabibbo favoured, but the ease of reconstruction of the final states for these modes results in tighter possible upper limits for the branching fractions and in addition the phase space enhancement helps in obtaining tighter exclusion curves for the mixing elements. Finally in Sec.~\ref{V}, we conclude.     

\section{Formalism for heavy neutrino mixing}
\label{2}


We extend the SM to include n right-handed SM singlets along with the three generation of left-handed SM SU(2)
doublets~\cite{tao han}:
  $$
                                         L_{aL}=\left(
                                         \begin{array}{c}
                                         \nu_{a}\\
                                          \ell_{a}\\
                                          \end{array}
                                         \right)_{L},\;\;\;N_{bR},
                                         $$
where a=1,2,3 and b=1,2,3,...,n. In this model, flavor eigenstates $\nu_{\ell L}$ can be written in terms of the mass eigenstates as, 
\begin{equation}\label{mixing relation}
\nu_{\ell L}=\sum_{m=1}^{3}U_{\ell m}\nu_{m L}+\sum_{m'=4}^{3+n}V_{\ell m'}N_{m' L}^{c},\,\,\,\text{with}\,\,UU^{\dagger}+VV^{\dagger}=1.
 \end{equation}
We take a phenomenological approach regarding the mass and mixing elements of the heavy singlet neutrino, taking them to be free parameters,
constrained only by experimental observations.
We denote by $V_{\ell N}$ the mixing coefficient between the standard flavour neutrino
$\nu_{\ell}$($\ell=e,\mu,\tau$) and the heavy mass eigenstate N. 
The charged current and neutral current interactions of the leptons in the basis of mass eigenstates are given by:
\begin{equation}\label{interaction lagrangian}
\begin{split}
&\mathcal{L}_{\ell}^{CC}=-\frac{g}{\sqrt{2}}W_{\mu}^{+}\left(\sum_{\ell =e}^{\tau}\sum_{m=1}^{3}U_{\ell m}^{*}\bar{\nu}_{m}\gamma^{\mu}P_{L}\ell+
\sum_{\ell =e}^{\tau}\sum_{m'=4}^{3+n}V_{\ell m'}^{*}\overline{N_{m'}^{c}}\gamma^{\mu}P_{L}\ell\right)+h. c,\\
&\mathcal{L}_{\ell}^{NC}=-\frac{g}{2cos\theta_{W}}Z_{\mu}\left(\sum_{\ell =e}^{\tau}\sum_{m=1}^{3}U_{\ell m}^{*}\bar{\nu}_{m}\gamma^{\mu}P_{L}\nu_{\ell}+
\sum_{\ell =e}^{\tau}\sum_{m'=4}^{3+n}V_{\ell m'}^{*}\overline{N_{m'}^{c}}\gamma^{\mu}P_{L}\nu_{\ell}\right)+h. c.,
\end{split}
 \end{equation}
 where $P_{L}=\frac{\left(1-\gamma_{5}\right)}{2}$, $\psi^{c}$ is the charged conjugate, $g$ is the $SU(2)_{L}$ gauge coupling. 
 The diagonalized majorana mass terms for the neutrinos can be written as:
 \begin{equation}
\mathcal{L}_{m}^{\nu}=-\frac{1}{2}\left(\sum_{m=1}^{3}m_{m}^{\nu}\overline{\nu_{m L}}\nu_{m R}^{c}+\sum_{m'=4}^{3+n}m_{m'}^{N}\overline{N_{m' L}^{c}}N_{m'R}\right)+h.c,
 \end{equation}

\section{$B_{c}^{-}\rightarrow \overline{B}_{s}^{0}\ell_{1}^{-}\ell_{2}^{-}\pi^{+}$ Decays}
\label{III} 
\subsection{Evaluation of the four-body decay rate}
\begin{figure}[H]
\subfigure[]{\includegraphics[width=7cm]{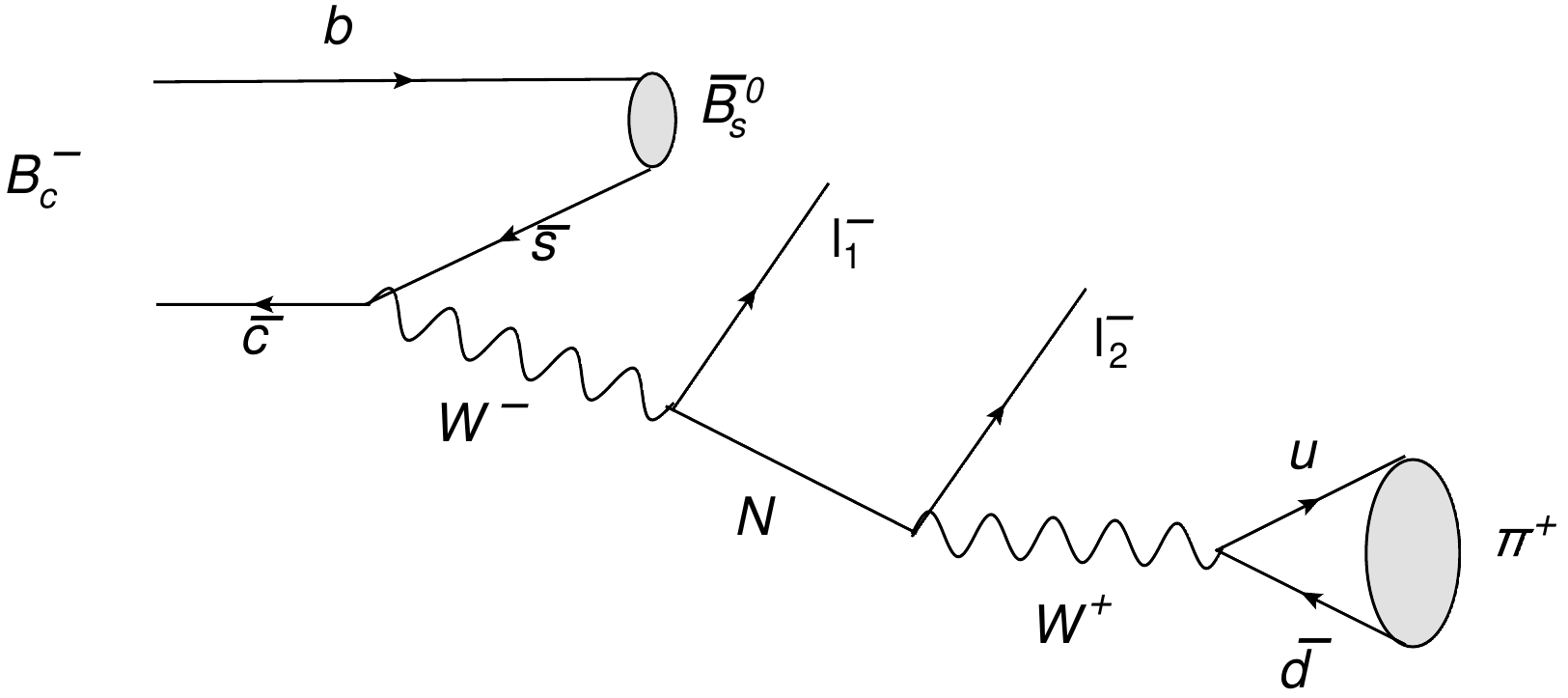}}
\subfigure[]{\includegraphics[width=7cm]{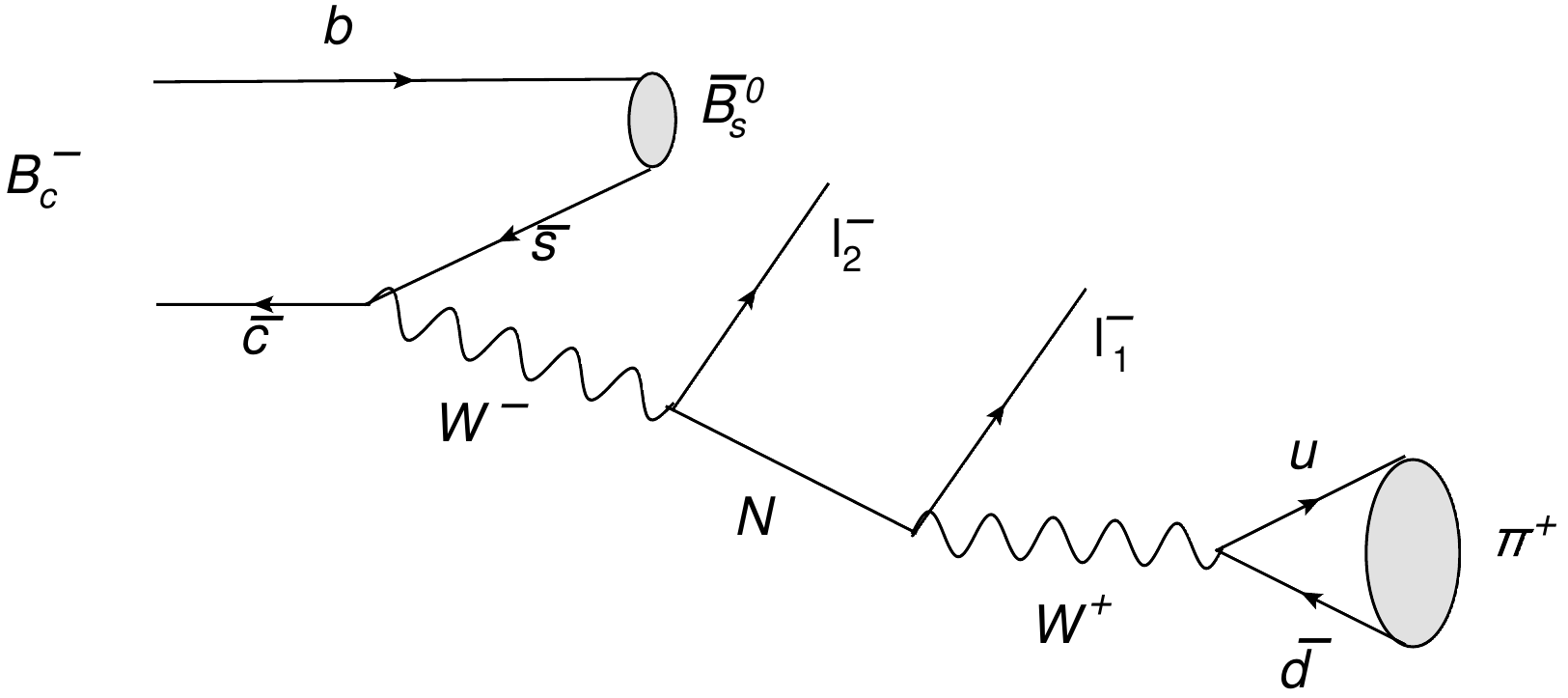}}
\caption{\small{Feynman diagrams for the decay $B_{c}^{-}\rightarrow \overline{B}_{s}^{0}\ell_{1}^{-}\ell_{2}^{-}\pi^{+}$}.}
\label{feynmandiagram}
\end{figure}

For the four-body decay $B_{c}^{-}(p)\rightarrow \overline{B}_{s}^{0}(k_{1})\ell_{1}(k_{2})\ell_{2}(k_{3})\pi^{+}(k_{4})$, where $\ell_{1},\ell_{2}=e,\mu$,
only s-channel diagrams shown in Fig.\ref{feynmandiagram} contribute. Hence, the Majorana neutrino N that induces this LNV process can appear as an intermediate on
mass shell state, leading to an enhancement of the decay rate. Note that the second diagram(Fig.\ref{feynmandiagram}(b)) arises from the exchange of the two leptons. We assume that
there is only one Majorana neutrino, that lies in the range, between $\sim (0.1-0.9)\,\text{GeV}$ that kinematically allows it to be on mass shell.
Moreover, being much heavier than the active light neutrinos, the cosmological and LEP bounds would imply that such a neutrino would have to be necessarily an electroweak gauge
singlet or sterile.

The decay amplitude for the processes depicted in Fig.\ref{feynmandiagram} can be expressed as,
\begin{equation}
 i\mathcal{M}=\left(\mathcal{M}_{lep}\right)_{\beta\mu}\left(\mathcal{M}_{had}\right)^{\beta\mu},
\end{equation}
where we can write the leptonic part as,
\begin{equation}
\label{leptonic}
\left(\mathcal{M}_{lep}\right)_{\beta\mu}=\frac{\sqrt{2}G_{F}V_{\ell_{1}N}^{*}V_{\ell_{2}N}m_{N}}{\left(p-k_{1}-k_{2}\right)^{2}-m_{N}^{2}+im_{N}\Gamma_{N}}
\bar{u}(k_{3})\gamma_{\beta}\gamma_{\mu}P_{R}v(k_{2})+\left(k_{2}\leftrightarrow k_{3},\ell_{1}\leftrightarrow \ell_{2}\right),
\end{equation}
where $G_{F}$ is the Fermi coupling constant,  $V_{\ell_{i} N}$(i=1, 2) are the mixing elements between the neutrino of flavour state $\nu_{\ell_{i}}$ and
mass eigenstate N and $\Gamma_{N}$ is the total decay width of the heavy neutrino N, obtained by summing over all accessible final states. The hadronic tensor is a product
of a transition matrix element of
$B_{c}$ to $B_{s}$,
and a matrix element for the production of a pion:
\begin{equation}
\left(\mathcal{M}_{had}\right)^{\beta\mu}=\frac{G_{F}}{\sqrt{2}}V_{cs}V_{ud}\braket{\overline{B}_{s}^{0}\left(k_{1}\right)|\bar{s}\gamma^{\mu}c|B_{c}^{-}\left(p\right)}
\braket{\pi^{+}\left(k_{4}\right)|\bar{u}\gamma^{\beta}d|0}~,
\end{equation}
where $V_{cs},\,V_{ud}$ are the Cabibbo-Kobayashi-Maskawa (CKM) matrix elements. The above two hadronic matrix elements can be written as,
\begin{equation}
\begin{split}
&\braket{\overline{B}_{s}^{0}\left(k_{1}\right)|\bar{s}\gamma^{\mu}c|B_{c}^{-}\left(p\right)}=\left(F_{+}(q^{2})(p+k_{1})^{\mu}+F_{-}(q^{2})(p-k_{1})^{\mu}\right),\\
&\braket{\pi^{+}\left(k_{4}\right)|\bar{u}\gamma^{\beta}d|0}=if_{\pi}k_{4}^{\beta},
\end{split}
\end{equation}
where $F_{+}\left(q^{2}\right),\,F_{-}\left(q^{2}\right)$ ($q\equiv p-k_{1}$) are the momentum transfer squared dependent $B_{c}^{-}$ to $\overline{B}_{s}^{0}$
transition form factors and $f_{\pi}$ is the decay constant of pion.
In terms of these form factors and decay constant, we can write the  amplitude
$\mathcal{M}$ as,
\begin{equation}\label{amplitude1}
\begin{split}
&\mathcal{M}=\frac{G_{F}^{2}V_{cs}V_{ud}V_{\ell_{1}N}^{*}V_{\ell_{2}N}f_{\pi}}{\left(p-k_{1}-k_{2}\right)^{2}-m_{N}^{2}+im_{N}\Gamma_{N}}
\left(F_{+}(q^{2})(p+k_{1})^{\mu}+F_{-}(q^{2})(p-k_{1})^{\mu}\right)\\
&\bar{u}(k_{3})\gamma_{\beta}\gamma_{\mu}\left(1+\gamma_{5}\right)v(k_{2})k_{4}^{\beta}+\left(k_{2}\rightarrow k_{3},\ell_{1}\leftrightarrow \ell_{2}\right)~.
\end{split}
\end{equation}

The form factors for $B_{c}^{-}\rightarrow\overline{B}_{s}^{0}$ have been calculated in the framework of 3-point QCD sum rule in Ref.~\cite{form factors}.
The $q^{2}$ dependence takes a simple pole form:
\begin{equation}
 F_{+}\left(q^{2}\right)=\frac{F_{+}(0)}{1-\frac{q^{2}}{M_{p}^{2}}}\,,\,\,\,\,\,F_{-}\left(q^{2}\right)=\frac{F_{-}(0)}{1-\frac{q^{2}}{M_{p}^{2}}},
\end{equation}
where $F_{+}(0)=1.3$ and $F_{-}(0)=-5.8$, and $M_{p}=1.7\div 1.8$ GeV. The accuracy of the sum rules used is determined by the variation of various parameters.
It is claimed in~\cite{form factors}
that these variations result in $\frac{\delta F}{F}\,\simeq\,5\%$. To avoid this theoretical uncertainty and model dependence in the form factors, we recommend
that the form factors should be determined
experimentally by measurement of the semileptonic mode, $B_{c}^{-}\rightarrow\overline{B}_{s}^{0}\mu^{-}\overline{\nu}_{\mu}$. Alternately, perhaps lattice
estimation of the form factors may also be possible.

Although the heavy sterile neutrino N is a SM singlet, it can decay via charged current and neutral current interactions, due to its mixing with the active neutrinos as is evident
from the Lagrangian~(\ref{interaction lagrangian}). 
The total decay width $\Gamma_{N}$ is given by:
\begin{equation}
\begin{split}
 &\Gamma_{N}=\sum_{\ell',P^{0}}\Gamma^{\nu_{\ell'}P^{0}}+\sum_{\ell',V^{0}}\Gamma^{\nu_{\ell'}V^{0}}+\sum_{\ell,P}2\Gamma^{\ell^{-} P^{+}}
 +\sum_{\ell,V}2\Gamma^{\ell^{-} V^{+}}\\
& +\sum_{\bar{\ell}_{1},\bar{\ell}_{2}(\bar{\ell}_{1}\neq\bar{\ell}_{2})}2\Gamma^{\bar{\ell}_{1}\bar{\ell}_{2}\nu_{\bar{\ell}_{2}}}+\sum_{\ell',\ell'_{2}}\Gamma^{\nu_{\ell'}\ell'_{2}\ell'_{2}}
+\sum_{\ell'}\Gamma^{\nu_{\ell'}\nu\bar{\nu}}.
 \end{split}
\end{equation}
In the mass range, which permits the heavy neutrino to be resonantly produced in the decay mode $B_{c}^{-}\rightarrow \overline{B}_{s}^{0}\ell_{1}^{-}\ell_{2}^{-}\pi^{+}$
the leptons $\ell,\,\bar{\ell}_{1},\,\bar{\ell}_{2},\,\ell'_{2}$ can be $e$ or $\mu$, while $\ell'$ can be $e, \mu$ or $\tau$, charged pseudoscalars ($P^{+}$) that can contribute are
$\pi^{+}$ and $K^{+}$, while $\pi^{0}$ and $\eta$ are the contributing neutral pseudoscalars ($P^{0}$), the charged vector mesons ($V^{+}$) will include $\rho^{+}$ and $K^{*+}$ and the neutral
vector mesons ($V^{0}$) that need to be included are $\rho^{0}$ and $\omega$\footnote{Note the $V^{0}$ cannot be $K^{*0}$ (or any other open flavour neutral meson), as the  $\nu_{\ell^{'}}V^{0}$ arises from a
	NC interaction, $K^{*0}$ can then only be produced via a flavour changing neutral current, which is not possible at tree level. We differ on this point from Refs.~\cite{GammaN calculation,tao han}.}.
	The detailed expressions for the decay rates for each of these channels can be found in Ref.~\cite{GammaN calculation,tao han}.
	
For the case of $B_{c}^{-}\rightarrow J/\psi\ell_{1}^{-}\ell_{2}^{-}\pi^{+}$ allowed mass range of $m_{N}$ is $(0.1-3)$ GeV. This will allow the additional charged pseudoscalar mesons: $D^{+}$, $D_{s}^{+}$ and charged vector mesons: $D^{*+}$, $D_{s}^{*+}$ to contribute, provided $\ell$ is either $e$ or $\mu$; for $\ell =\tau$ the mesons can only be $\pi^{+}$, $K^{+}$, $\rho^{+}$, $K^{*+}$. Additional contributing neutral pseudoscalar mesons are: $\eta'$ and $\eta_{c}$ while, $\phi$ and $J/\psi$   are the heavier neutral vector mesons that can also be produced in the decays of N. $\bar{\ell}_{1}$ or $\bar{\ell}_{2}$ can now also be a $\tau$.

For the case of $B_{c}^{-}\rightarrow\ell_{1}^{-}\ell_{2}^{-}\pi^{+}$ the allowed mass range in which N can be resonantly produced is $(0.1-6)\,\text{GeV}$.
Charged pseudoscalar meson $B^{+}$ and vector meson $B^{*+}$
will also contribute now for $\ell=e, \mu$.
For $\ell=\tau$, the additional accompanying mesons will be $D^{+}$, $D_{s}^{+}$, $D^{*+}$, $D_{s}^{*+}$. Also, $\ell'_{2}$ can also be $\tau$.

\begin{figure}[H]
\centering
\includegraphics[width=0.7\textwidth]{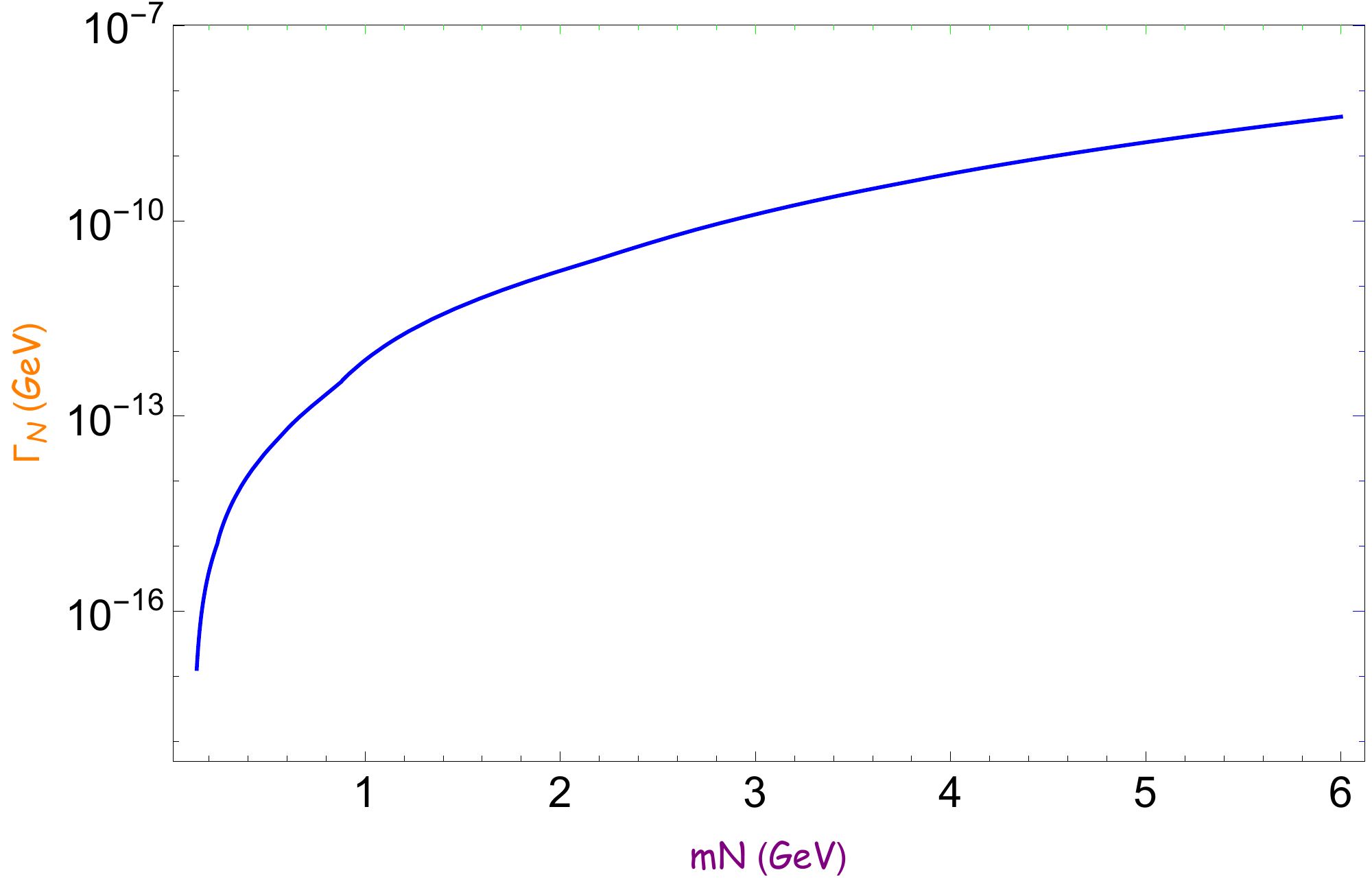}
\caption{\small{Heavy neutrino decay width, $\Gamma_{N}$ as a function of the mass $m_{N}$ when the magnitude of all the mixing angles $\mid V_{\ell N}\mid=1$ ($\ell=e, \mu, \tau$). A bigger range for $m_{N}$ is chosen than that which allows a resonant enhancement of the $B_{c}^{-}\rightarrow \overline{B}_{s}^{0}\ell_{1}^{-}\ell_{2}^{-}\pi^{+}$ decay, so as to include the larger values of $m_N$ that will be permitted by the other $B_c$ decay modes to be discussed in Sec.~\ref{IV}}.}
\label{gammaNplot}
\end{figure}

\begin{figure}[H]
\centering
\includegraphics[width=0.7\textwidth]{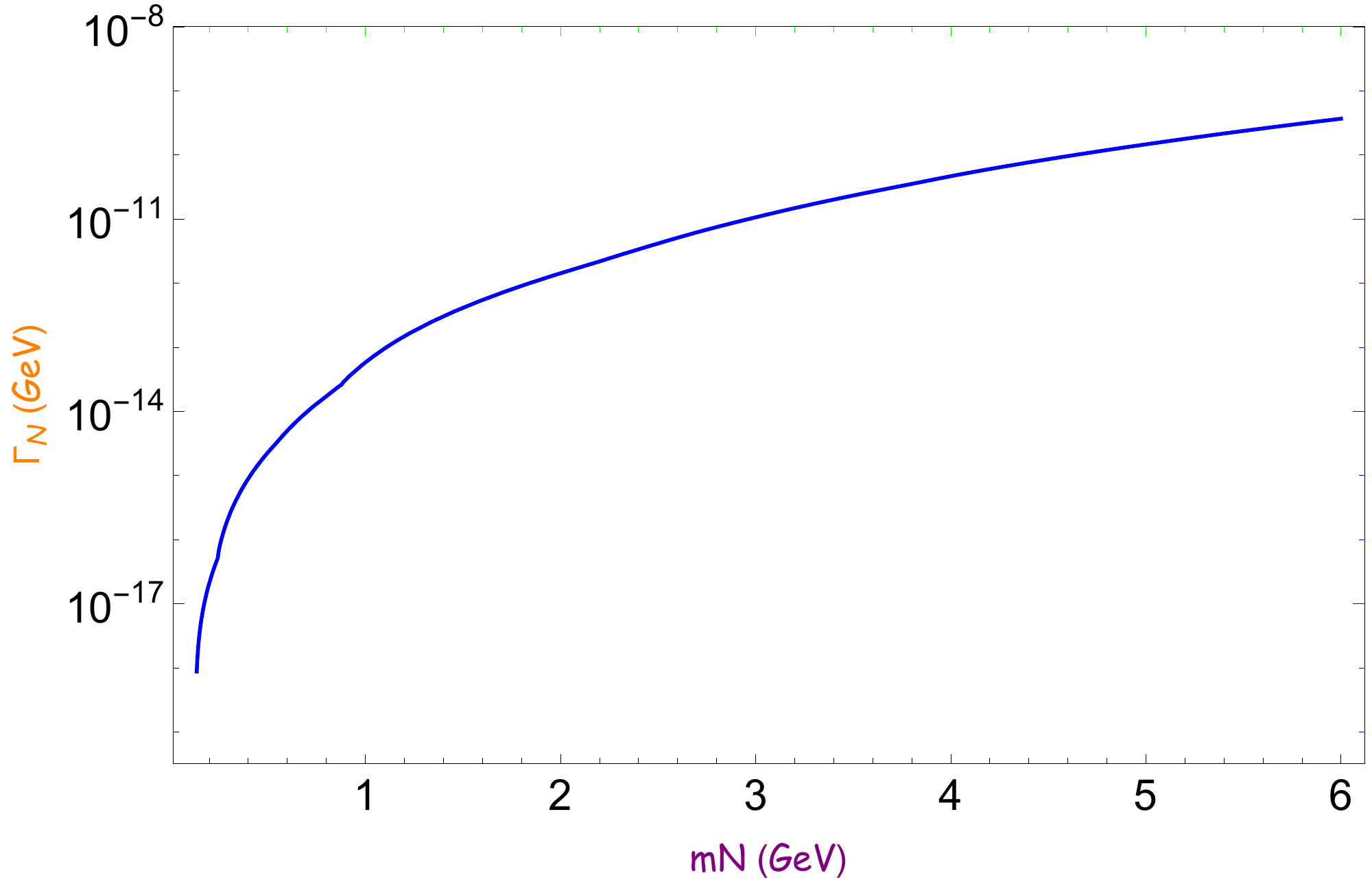}
\caption{\small{Heavy neutrino decay width as a function of the mass $m_{N}$ with the maximum values of the mixing angles, $\mid V_{\ell N}\mid^{2}, \ell=e, \mu, \tau$ allowed by unitarity and the Global fits to oscillation data}.}
\label{GammaNplot2}
\end{figure}

We have re-evaluated $\Gamma_{N}$ using the meson masses and decay constants from Ref.~\cite{PDG}, in the relevant mass range for the $B_{c}$ decay modes considered here and write it in the form,
\begin{equation}
 \Gamma_{N}=a_{e}\left(m_{N}\right)\mid V_{eN}\mid^{2}+a_{\mu}\left(m_{N}\right)\mid V_{\mu N}\mid^{2}+a_{\tau}\left(m_{N}\right)\mid V_{\tau N}\mid^{2}~,
\end{equation}
where, $a_e$, $a_\mu$ and $a_\tau$ are functions of the Majorana neutrino mass and hence will differ from mode to mode. In Fig.~\ref{gammaNplot}, we plot the decay width $\Gamma_{N}$ as function of mass
$m_{N}$, for the mixings $\mid V_{eN}\mid=\mid V_{\mu N}\mid=\mid V_{\tau N}\mid=1$. 
The unitarity condition in eqn.(\ref{mixing relation}) implies the following constraints on the mixing elements, $\mid V_{eN}\mid^{2}$, $\mid V_{\mu N}\mid^{2}$ and
$\mid V_{\tau N}\mid^{2}$:
\begin{equation}\label{unitarity condition}
	\begin{split}
		&\mid U_{e1}\mid^{2}+\mid U_{e2}\mid^{2}+\mid U_{e3}\mid^{2}+\mid V_{eN}\mid^{2}=1,\\
		&\mid U_{\mu 1}\mid^{2}+\mid U_{\mu 2}\mid^{2}+\mid U_{\mu 3}\mid^{2}+\mid V_{\mu N}\mid^{2}=1,\\
		&\mid U_{\tau 1}\mid^{2}+\mid U_{\tau 2}\mid^{2}+\mid U_{\tau 3}\mid^{2}+\mid V_{\tau N}\mid^{2}=1,\\
	\end{split}
\end{equation}
where $U_{ei}$, $U_{\mu i}$, and $U_{\tau i}$, i=1,2,3 are the PMNS matrix elements. Using the 3$\sigma$ ranges of the PMNS matrix elements obtained from a Global
analysis of neutrino oscillation data~\cite{Global Analyses of Neutrino Oscillation Experiments}, we calculate the bounds on $\mid V_{eN}\mid^{2},\,\mid V_{\mu N}\mid^{2},\,\mid V_{\tau N}\mid^{2}$ to be:
\begin{equation}\label{oscillation bound}
	\mid V_{eN}\mid^{2}\leq0.075434,\,\,\mid V_{\mu N}\mid^{2}\leq0.377898,\,\,\mid V_{\tau N}\mid^{2}\leq0.376088.
\end{equation}
The $3\sigma$ ranges of the PMNS matrix elements of Ref.~\cite{Global Analyses of Neutrino Oscillation Experiments} are consistent with those obtained by a study of unitarity of
the neutrino mixing matrix in~\cite{Unitarity and the three flavour neutrino mixing matrix}.
Fig.~\ref{GammaNplot2} displays the heavy neutrino decay width without any assumptions, and using the maximum values for $\mid V_{\ell N}\mid^{2}, \ell=e, \mu, \tau$ permitted by unitarity and global fits to neutrino oscillation data.
For the mass range of our interest, $\Gamma_{N}$ is very small, $\mathcal{O}\left(10^{-17}-10^{-8}\right)\,\text{GeV}$, if the mixing
$\mid V_{eN}\mid^{2}=\mid V_{\mu N}\mid^{2}=\mid V_{\tau N}\mid^{2}=1$ and even smaller for more realistic values of these mixing angles. Due to this narrow decay width of $N$, the two propagators for N, in equation (\ref{amplitude1}) can be written as,
\begin{equation}\label{narrow width defn}
\frac{1}{\left(p_{N}^{2}-m_{N}^{2}\right)^{2}+m_{N}^{2}\Gamma_{N}^{2}}\simeq\frac{\pi}{m_{N}\Gamma_{N}}\delta\left(p_{N}^{2}-m_{N}^{2}\right).
\end{equation}
Moreover, in the narrow width approximation the two channels contribute as a sum to the total decay width, as the interference term is neglegible.

Most of the earlier studies of LNV meson and tau decays have focused on three-body decays. A few more recent phenomenological studies~\cite{fourbody decay,four-body tau,Yuan, Dong,Lepton number violating decay} of four-body
LNV processes have also been performed, including an experimental search through the mode $B^-\rightarrow D^0\pi^+\mu^-\mu^-$ by LHCb~\cite{LHCb2012}. The particular four-body $B_c$ decay mode being considered here has the advantage of being Cabibbo favored and hence enhanced.  

To calculate the four-body phase space required for evaluating the decay rate
$\Gamma(B_{c}^{-}(p)\rightarrow\overline{B}_{s}^{0}(k_{1})\ell_{1}(k_{2})\ell_{2}(k_{3})\pi^{+}(k_{4}))=
\frac{1}{2m}\int d_{4}(ps)\mid\mathcal{M}\mid^{2}$, the final particles can be partitioned into two subsystems $X_{12}$ and $X_{34}$, each of which subsequently
decays into a two-body state. Hence, the four-body phase space integral is decomposed into a product of three two-body phase space integrals:
\begin{equation}
 d_{4}(ps)=d_{2}\left(ps\;B_{c}^{-}\rightarrow X_{12}X_{34}\right)d_{2}\left(ps\;X_{12}\rightarrow k_{1}k_{2}\right)d_{2}\left(ps\;X_{34}\rightarrow k_{3}k_{4}\right)
 dM_{12}^{2}dM_{34}^{2},
\end{equation}
where $X_{12}=(k_{1}+k_{2})$, $X_{34}=(k_{3}+k_{4})$, $X_{12}^{2}=M_{12}^{2}$ and $X_{34}^{2}=M_{34}^{2}$, $p^{2}=m^{2}$ and $k_{i}^{2}=m_{i}^{2}$.
The four-body phase space therefore takes the form,
\begin{equation}\label{phasespace}
\begin{split}
&d_{4}(ps)=\frac{1}{n!}\frac{1}{(4\pi)^{6}}\frac{1}{4}\lambda^{\frac{1}{2}}\left(1,\frac{M_{12}^{2}}{m^{2}},\frac{M_{34}^{2}}{m^{2}}\right)
\lambda^{\frac{1}{2}}\left(1,\frac{m_{1}^{2}}{M_{12}^{2}},\frac{m_{2}^{2}}{M_{12}^{2}}\right)\\
&\lambda^{\frac{1}{2}}\left(1,\frac{m_{3}^{2}}{M_{34}^{2}},\frac{m_{4}^{2}}{M_{34}^{2}}\right)dM_{12}^{2}dM_{34}^{2}dcos\theta_{12}dcos\theta_{34}d\phi,
 \end{split}
\end{equation}
where m, $m_{1}$, $m_{2}$, $m_{3}$, and $m_{4}$ are the masses of $B_{c}^{-}$, $\overline{B}_{s}^{0}$, $\ell_{1}$, $\ell_{2}$ and $\pi^{+}$ respectively, $\lambda(x,y,z)=
x^{2}+y^{2}+z^{2}-2xy-2xz-2yz$, and n=2 for identical leptons in the final state, otherwise n=1. $\theta_{12}$($\theta_{34}$) is the angle in the $\vec{X_{12}}$($\vec{X_{34}}$) rest frame between the three momentum
$\vec{k_{1}}$($\vec{k_{3}}$)  and the line of flight of $\vec{X_{12}}$($\vec{X_{34}}$) in the $B_{c}$ rest frame. The angle $\phi$ is the angle between
the normals to the planes defined in the $B_{c}$ rest frame by the $\overline{B}_{s}^{0}\ell_{1}$ pair and the $\ell_{2}\pi^{+}$ pair. This is depicted in the four-body kinematics diagram in the appendix.  The four momenta $k_{1}, k_{2}$ ($k_{3}, k_{4}$) are first evaluated in the $\vec{X_{12}}$($\vec{X_{34}}$) rest frame. To finally evaluate the decay rate in the $B_{c}$ rest frame, it is assumed that $\vec{X_{12}}$ moves in the +$\hat{z}$ direction and $\vec{X_{34}}$ in the -$\hat{z}$ direction and the resultant boosted explicit form of all the four momenta in the $B_{c}^{-}$ rest frame are also given in the appendix.  

\begin{figure}[H]
\subfigure[]{\includegraphics[width=8cm]{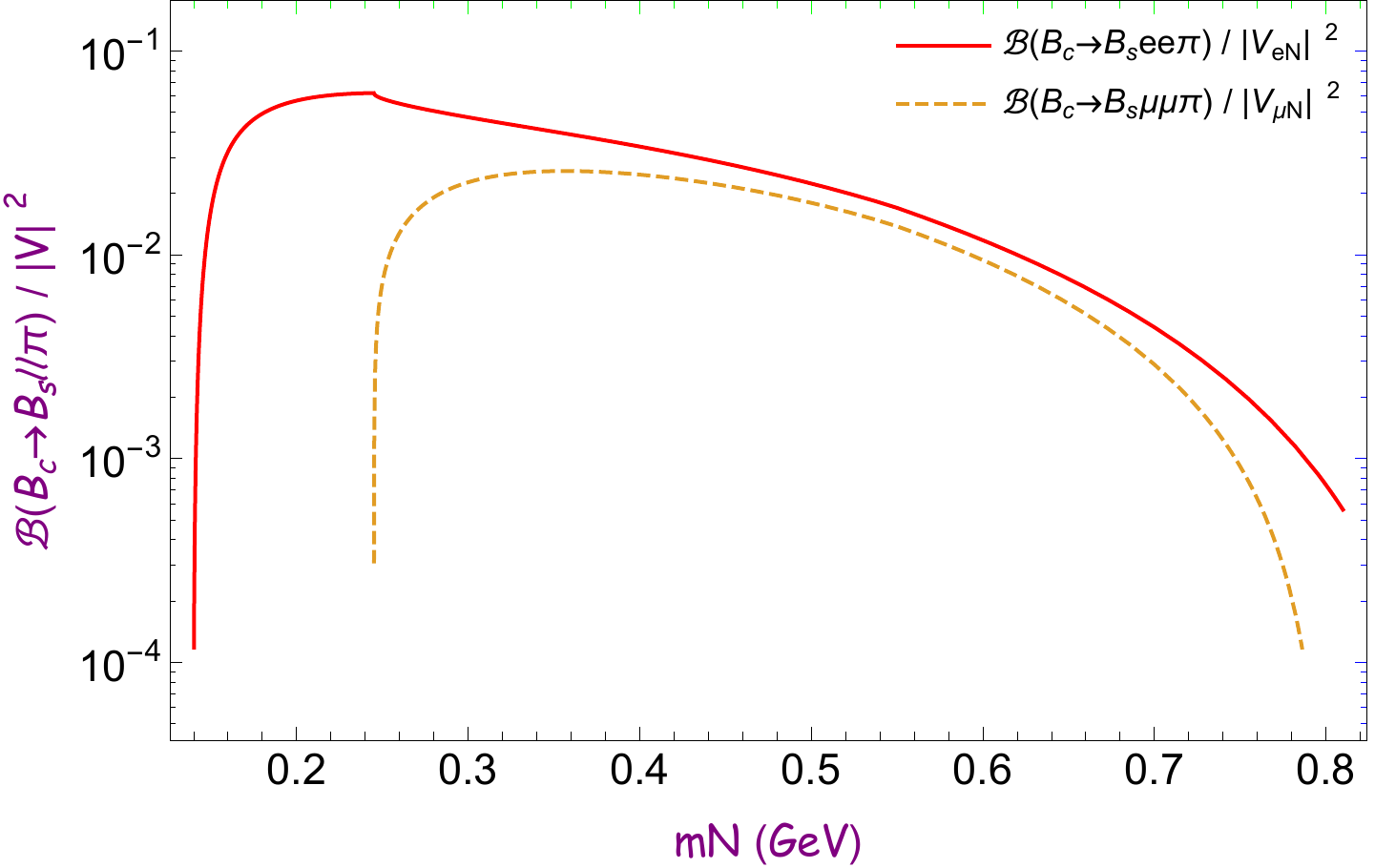}}
\subfigure[]{\includegraphics[width=8cm]{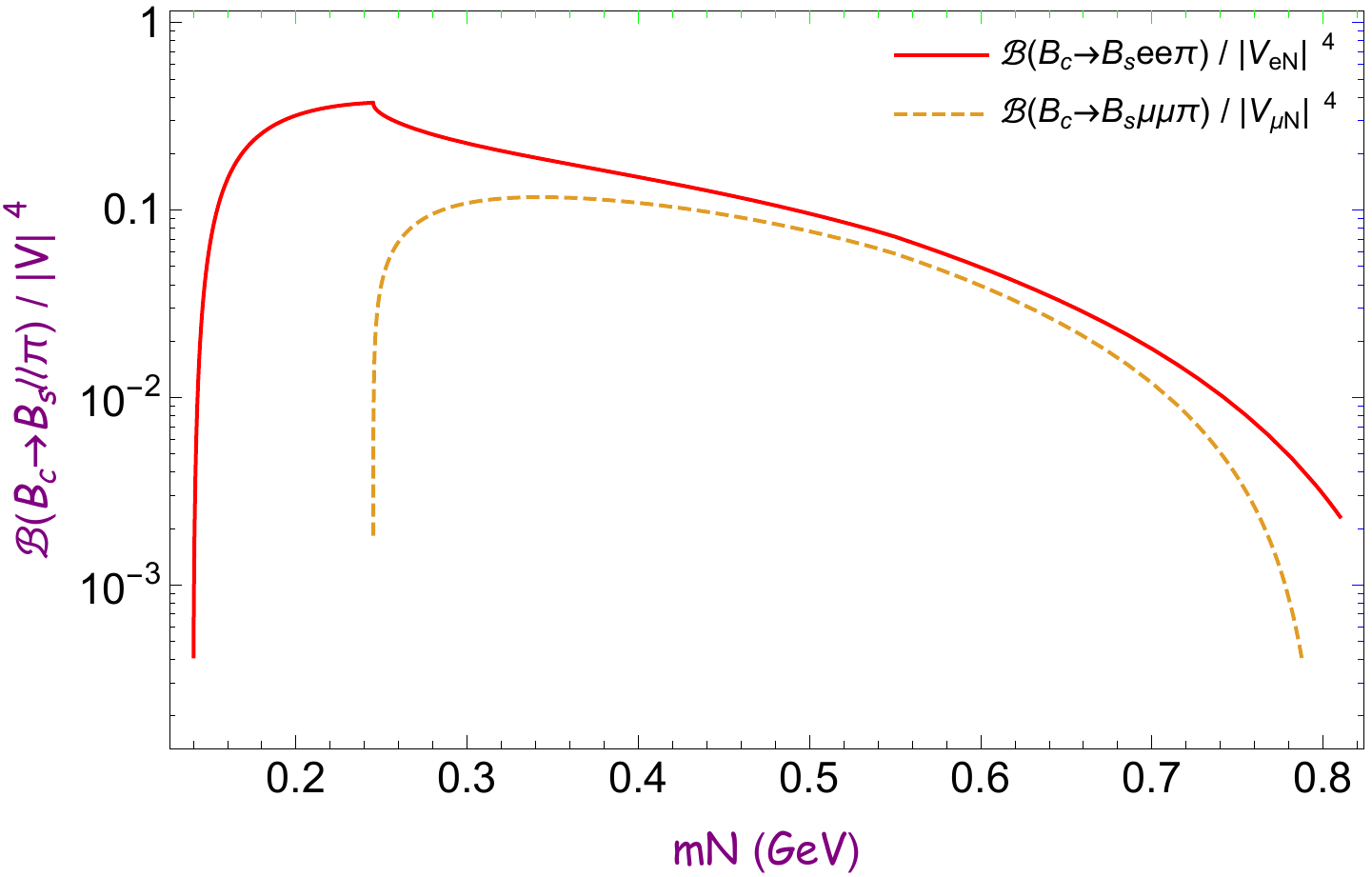}}
\caption{\small{$\frac{\mathscr{B} \left(B_{c}^{-}\rightarrow \overline{B}_{s}^{0}\ell^{-}\ell^{-}\pi^{+}\right)}{\mid V_{\ell N}\mid^{2}}$/
$\frac{\mathscr{B} \left(B_{c}^{-}\rightarrow \overline{B}_{s}^{0}\ell^{-}\ell^{-}\pi^{+}\right)}{\mid V_{\ell N}\mid^{4}}$,
where, $\ell= e,\mu$. The theoretical calculation uses $\Gamma_{N}$ obtained (a)with the assumption $\mid V_{eN}\mid\,\sim\,\mid V_{\mu N}\mid\,\sim\,\mid V_{\tau N}\mid$ (shown on the left), (b)using the upper limits of the mixing elements allowed by unitarity and Global fits to oscillation data (shown on the right).}}
\label{BR}
\end{figure}

Alternately, rather than calculating the full 4-body kinematics to evaluate the decay rate, the narrow width approximation can be used to
evaluate the decay rate as a product of a 3-body decay rate and the branching ratio for decay of N to a 2-body mode,  as specified below:
\begin{equation}\label{alternative definition of decay rate}
\begin{split}
&\Gamma\left(B_{c}^{-}\rightarrow \overline{B}_{s}^{0}\ell_{1}^{-}\ell_{2}^{-}\pi^{+}\right)\approx\Gamma\left(B_{c}^{-}\rightarrow \overline{B}_{s}^{0}\ell_{1}^{-}N\right).
\frac{\Gamma\left(N\rightarrow \ell_{2}^{-}\pi^{+}\right)}{\Gamma_{N}}.\\
\end{split}
\end{equation}
In Fig.~\ref{BR}, we show the curves corresponding to $\frac{\mathscr{B} \left(B_{c}^{-}\rightarrow \overline{B}_{s}^{0}e^{-}e^{-}\pi^{+}\right)}{\mid V_{e N}\mid^{2}}$/
$\frac{\mathscr{B} \left(B_{c}^{-}\rightarrow \overline{B}_{s}^{0}e^{-}e^{-}\pi^{+}\right)}{\mid V_{e N}\mid^{4}}$ and
$\frac{\mathscr{B}\left(B_{c}^{-}\rightarrow \overline{B}_{s}^{0}\mu^{-}\mu^{-}\pi^{+}\right)}{\mid V_{\mu N}\mid^{2}}$/
$\frac{\mathscr{B}\left(B_{c}^{-}\rightarrow \overline{B}_{s}^{0}\mu^{-}\mu^{-}\pi^{+}\right)}{\mid V_{\mu N}\mid^{4}}$, as a function of the heavy neutrino mass, $m_N$. The regions below the curves are theoretically allowed. For this calculation, $\Gamma_{N}$ is  evaluated, either under the assumption that has been frequently used in the literature~\cite{tao han,GammaN calculation}, $\mid V_{eN}\mid\,\sim\,\mid V_{\mu N}\mid\,\sim\,\mid V_{\tau N}\mid$, shown in the left figure (a) or, using the upper limits of the mixing elements, obtained in
eqn.(\ref{oscillation bound}), leading to the maximum value of $\Gamma_{N}$ permitted by unitarity and Global fits to oscillation data, displayed in the right figure (b).
Note that the few kinks  in the plots in Figs.~(\ref{gammaNplot}-\ref{BR}) arise from threshold for a new channel contributing to $\Gamma_N$ at the corresponding $m_N$ value (e.g. around $0.135\,\text{GeV}$ and $0.245\,\text{GeV}$ the visible kinks are from the threshold for $\nu\pi^0$ and $\pi^+\mu^-$ respectively).
\subsection{Bounds on Mixing angles using upper limits on the Branching ratios for $B_{c}^{-}\rightarrow \overline{B}_{s}^{0}\ell_{1}^{-}\ell_{2}^{-}\pi^{+}$ Decays}
Using the matrix element in eqn.(\ref{amplitude1}) and the narrow width approximation, eqn.(\ref{narrow width defn}) the LNV branching ratios can be written as:
\begin{equation}\label{analytical1}
\begin{split}
&\mathscr{B} \left(B_{c}^{-}\rightarrow \overline{B}_{s}^{0}e^{-}e^{-}\pi^{+}\right)=G_{ee}\left(m_{N}\right)\frac{\mid V_{e N}\mid^{4}}{\Gamma_{N}}~,\\
&\mathscr{B} \left(B_{c}^{-}\rightarrow \overline{B}_{s}^{0}\mu^{-}\mu^{-}\pi^{+}\right)=G_{\mu\mu}\left(m_{N}\right)\frac{\mid V_{\mu N}\mid^{4}}{\Gamma_{N}}~,
\end{split}
\end{equation}
where, $G_{ee}$ and $G_{\mu\mu}$ are functions of the Majorana mass and depend on the explicit matrix element and phase space for each of the processes.
When both the like sign dileptons in Fig.\ref{feynmandiagram} are not of the same flavour, then the process is not only lepton number violating but also lepton flavour violating.
If the two vertices of N production and decay can be separated, then the two processes, $B_{c}^{-}\rightarrow \overline{B}_{s}^{0}e^{-}N$ followed by $N\rightarrow \mu^{-}\pi^{+}$
and $B_{c}^{-}\rightarrow \overline{B}_{s}^{0}\mu^{-}N$ followed by $N\rightarrow e^{-}\pi^{+}$ can be distinguished. Assuming this separation, we may write: 
\begin{equation}\label{analytical1p}
\begin{split}
&\mathscr{B} \left(B_{c}^{-}\rightarrow \overline{B}_{s}^{0}e^{-}\mu^{-}\pi^{+}\right)=G_{e\mu}\left(m_{N}\right)\frac{\mid V_{e N}\mid^{2}\mid V_{\mu N}\mid^{2}}{\Gamma_{N}}~,\\
&\mathscr{B} \left(B_{c}^{-}\rightarrow \overline{B}_{s}^{0}\mu^{-}e^{-}\pi^{+}\right)=G_{\mu e}\left(m_{N}\right)\frac{\mid V_{e N}\mid^{2}\mid V_{\mu N}\mid^{2}}{\Gamma_{N}}~,
\end{split}
\end{equation}
where, we use the notation that the first lepton is produced along with the $N$, while the second lepton is produced in the decay of $N$; $G_{e\mu}$ ($G_{\mu e}$) are again functions of the Majorana mass and vary with the explicit matrix element and phase space for each of the processes. 
Now, defining,
\begin{equation}\label{analytical2}
\begin{split}
&F_{ee}\equiv\frac{\mathscr{B}^{exp}\left(B_{c}^{-}\rightarrow \overline{B}_{s}^{0}e^{-}e^{-}\pi^{+}\right)}{G_{ee}\left(m_{N}\right)},\\
&F_{\mu\mu}\equiv\frac{\mathscr{B}^{exp}\left(B_{c}^{-}\rightarrow \overline{B}_{s}^{0}\mu^{-}\mu^{-}\pi^{+}\right)}{G_{\mu\mu}\left(m_{N}\right)},\\
&F_{e\mu}\equiv\frac{\mathscr{B}^{exp}\left(B_{c}^{-}\rightarrow \overline{B}_{s}^{0}e^{-}\mu^{-}\pi^{+}\right)}{G_{e\mu}\left(m_{N}\right)},\\
&F_{\mu e}\equiv\frac{\mathscr{B}^{exp}\left(B_{c}^{-}\rightarrow \overline{B}_{s}^{0}\mu^{-}e^{-}\pi^{+}\right)}{G_{\mu e}\left(m_{N}\right)},
\end{split}
\end{equation}
where $\mathscr{B}^{exp}$ are the expected experimental upper limits of the Branching ratios, we can obtain the constraints: 
\begin{equation}\label{analytical3}
\frac{\mid V_{e N}\mid^{4}}{\Gamma_{N}}\,<\,F_{ee}\,\,,\,\,\frac{\mid V_{\mu N}\mid^{4}}{\Gamma_{N}}\,<\,F_{\mu\mu},\,\,
\frac{\mid V_{e N}\mid^{2}\mid V_{\mu N}\mid^{2}}{\Gamma_{N}}\,<\,F_{e\mu}/F_{\mu e}~.
\end{equation}
The upper limits on the $\mathscr{B}^{exp}$ in eqn.(\ref{analytical3}) can be very simply translated into the upper limits on, $\mid V_{e N}\mid^{2}$, $\mid V_{\mu N}\mid^{2}$,
$\mid V_{e N}V_{\mu N}\mid$ under the assumption, 
$\mid V_{eN}\mid\,\sim\,\mid V_{\mu N}\mid\,\sim\,\mid V_{\tau N}\mid$ in $\Gamma_{N}$.
This leads eqn.(\ref{analytical3}) to result in the constraints,
\begin{equation}\label{analytical4}
\begin{split}
&\mid V_{e N}\mid^{2}<F_{ee}\left(a_{e}+a_{\mu}+a_{\tau}\right);\,\,\mid V_{\mu N}\mid^{2}<F_{\mu\mu}\left(a_{e}+a_{\mu}+a_{\tau}\right);\\
&\mid V_{eN}V_{\mu N}\mid<F_{e\mu}/F_{\mu e }\left(a_{e}+a_{\mu}+a_{\tau}\right).
\end{split}
\end{equation}
According to Ref.~\cite{Chang} at the LHC with $\sqrt{s}=14\,\text{TeV}$, the beam luminosity and production cross-section will be high enough that the rate of producing
$B_c$ events can be $10^{8}-10^{9}$ per year. A crude estimate~\cite{Belyaev} using the measured~\cite{BcProdLHCb} ratio of production cross section times branching fractions between
the $B_c^+\rightarrow J/\Psi\pi^+$ and $B^+\rightarrow J/\Psi K^+$ decays at $\sqrt{s}=8\,\text{TeV}$, indicates  $\sim \mathcal{O}\left(10^9-10^{10}\right)$ $B_c$ events with $10\,\text{fb}^{-1}$
luminosity at $13/14\,\text{TeV}$. Ultimately, the production cross-section will be directly measured by LHCb at $\sqrt{s}=13/14\,\text{TeV}$ and will be known more precisely. In any case the large
number of $B_c$ events will make a search for the proposed rare LNV $B_c$ decays feasible. Even if these decay modes are not seen, one may naively estimate that it may be possible to set upper limits on the branching ratios
of $\sim \mathcal{O}\left(10^{-7}-10^{-9}\right)$. However, since the final $B_s$ meson needs to be reconstructed via its prominant decay modes, either $B_s\rightarrow J/\psi(\mu\mu)\phi(KK)$ or $B_s\rightarrow D_s(KK\pi)\pi$,
with $\mathscr{B}(B_s\rightarrow J/\psi\phi)\times \mathscr{B}(J/\psi\rightarrow\mu\mu)\times \mathscr{B}(\phi\rightarrow KK)\sim \mathcal{O}(10^{-5})$;
$\mathscr{B}(B_s\rightarrow  D_s\pi\times \mathscr{B}(D_s\rightarrow KK\pi)\sim \mathcal{O}(10^{-4})$, upper limits on $\mathscr{B} \left(B_{c}^{-}\rightarrow \overline{B}_{s}^{0}\ell_{1}^{-}\ell_{2}^{-}\pi^{+}\right)$ of only $\sim \mathcal{O}(10^{-5}-10^{-4})$ may be feasible. These limits are just indicative, exact realistic limits will only be determined by the experimental collaboration, after incorporating the 
detection, reconstruction efficiencies of all the final particles. Of course, tighter limits would be possible at future colliders.

\begin{figure}
\subfigure[]{\includegraphics[width=8cm]{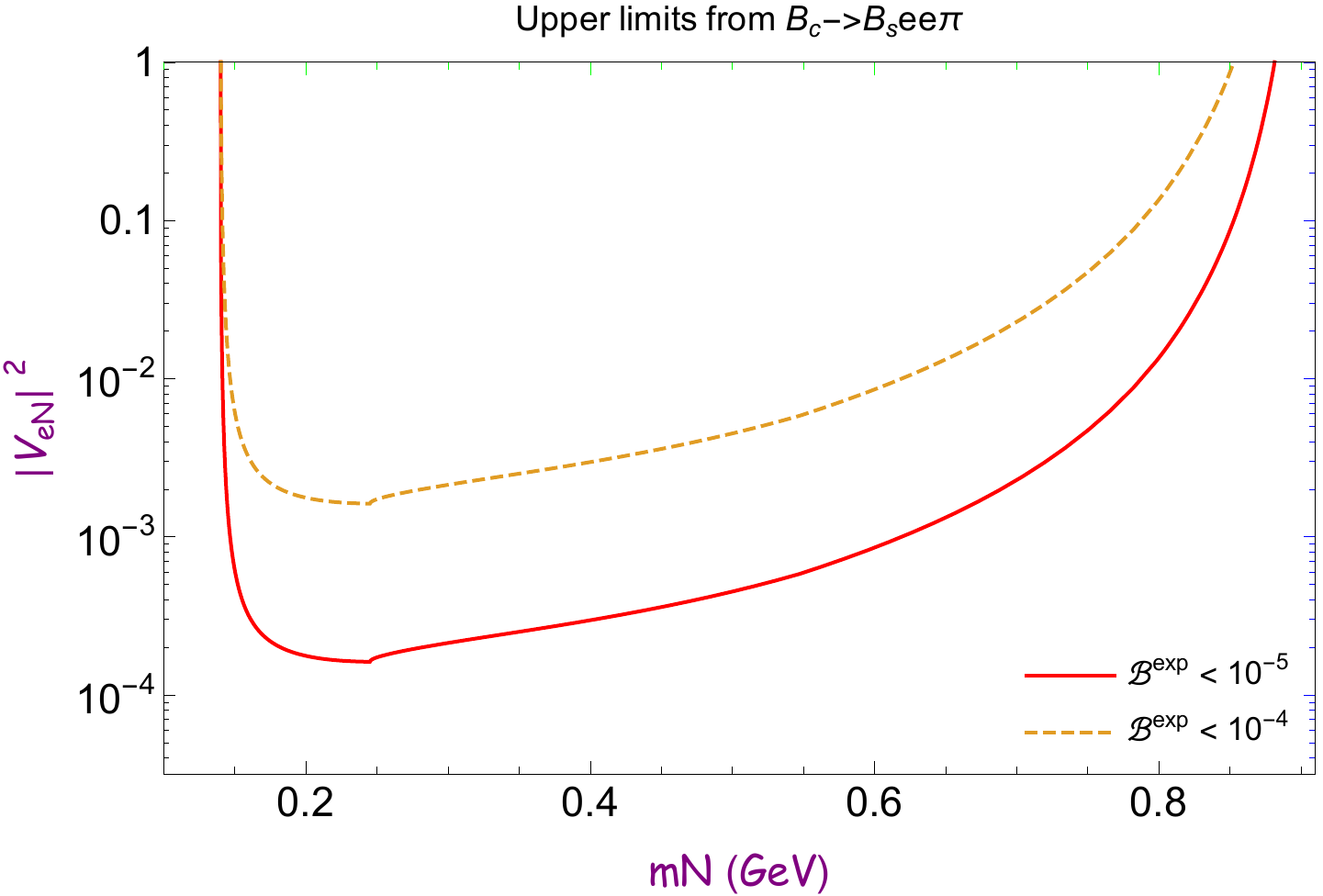}}
\subfigure[]{\includegraphics[width=8cm]{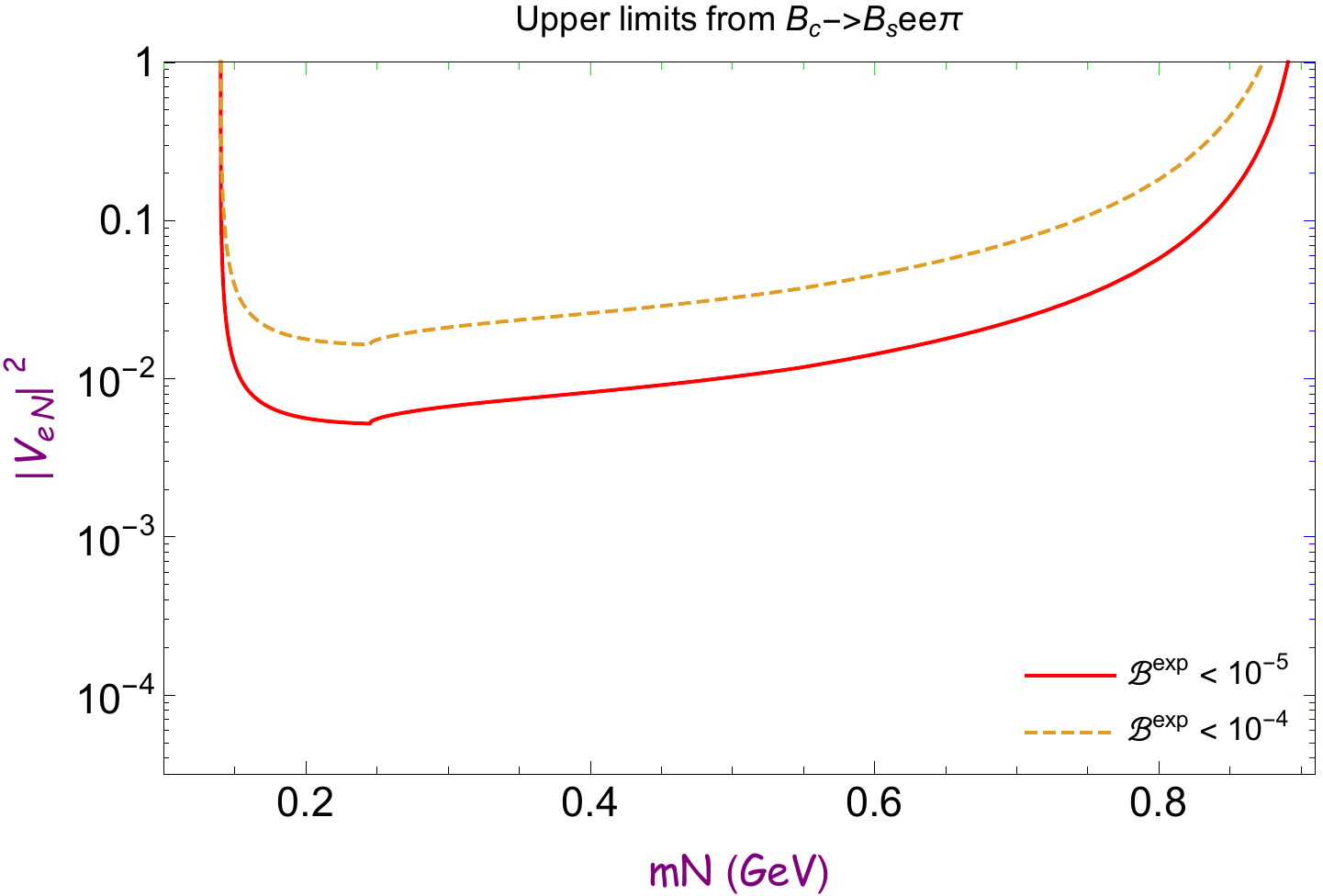}}
\caption{\small{Exclusion curves for the mixing element $\mid V_{eN}\mid^{2}$ corresponding to the different expected upper limits for branching ratio of the decay mode
$B_{c}^{-}\rightarrow \overline{B}_{s}^{0}e^{-}e^{-}\pi^{+}$. In $\Gamma_{N}$, the left figure(a) uses the assumption,
$\mid V_{eN}\mid\,\sim\,\mid V_{\mu N}\mid\,\sim\,\mid V_{\tau N}\mid$, while the right figure(b) uses the maximum allowed magnitude of the mixing elements
$V_{eN}$, $V_{\mu N}$, $V_{\tau N}$ from unitarity and the Global fits to oscillation data.}}
\label{1}
\end{figure}

\begin{figure}
\subfigure[]{\includegraphics[width=8cm]{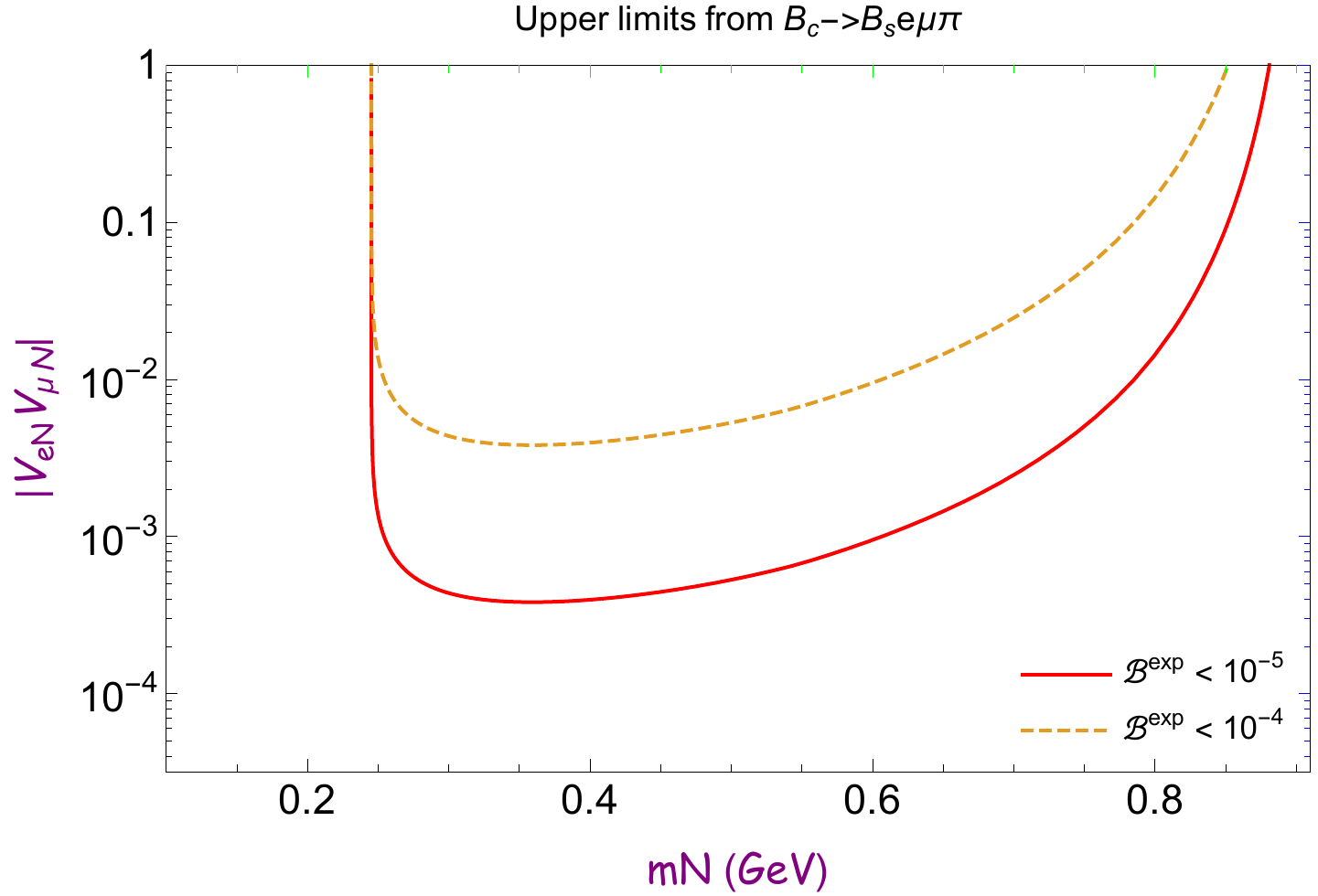}}
\subfigure[]{\includegraphics[width=8cm]{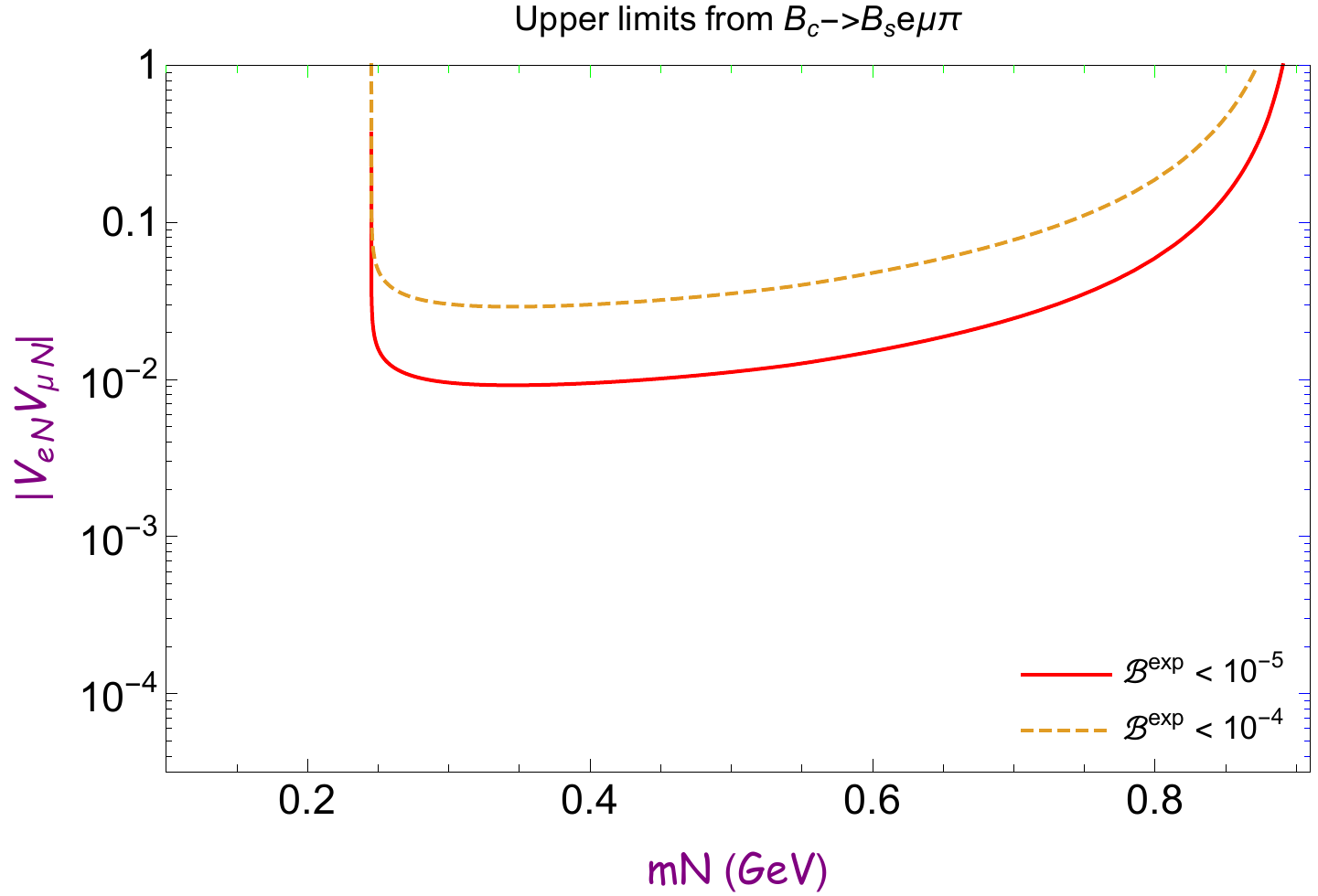}}
\subfigure[]{\includegraphics[width=8cm]{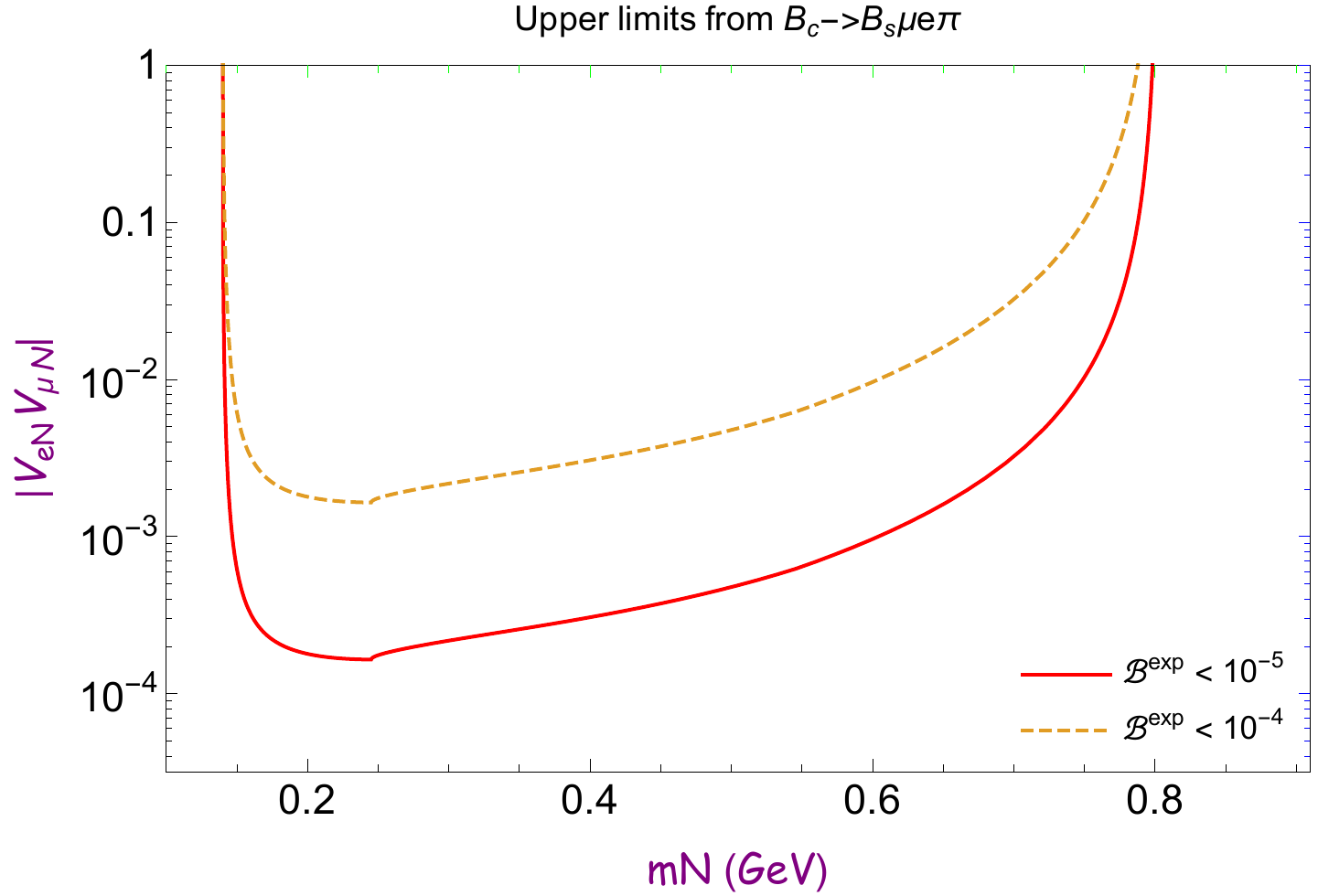}}
\subfigure[]{\includegraphics[width=8cm]{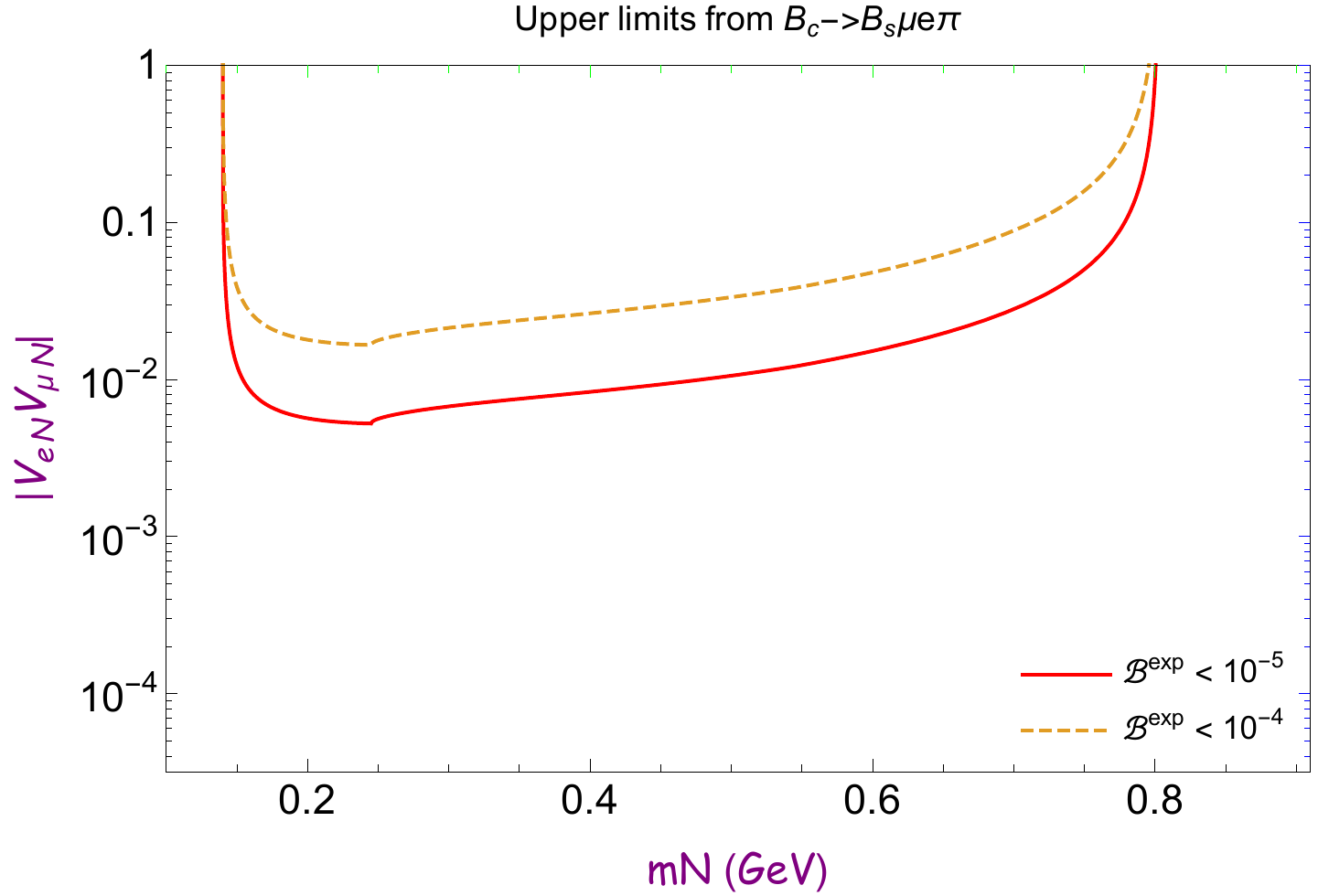}}
\caption{\small{Exclusion curves for the mixing element $\mid V_{\ell_1 N} V_{\ell_2 N}\mid$ from the possible upper limits for the branching fraction of the decay mode $B_{c}^{-}\rightarrow\overline{B}_{s}^{0}\ell_1^{-}\ell_2^{-}\pi^{+}$. The upper plots are for $\ell_1=e$ and $\ell_2=\mu$, while the lower plots are for $\ell_1=\mu$ and $\ell_2=e$. For $\Gamma_N$, the figures on the left use the assumption of equal magnitudes of all the mixing elements while those on
the right use the maximum values of the mixing elements permissible by the unitarity constraints and global fits to oscillation data.}}
\label{3}
\end{figure}

In the left panels of the Figs. \ref{1}, \ref{2} and \ref{3}, we show the exclusion curves corresponding to the constraints on the mixing angles
$\mid V_{eN}\mid^{2}$, $\mid V_{\mu N}\mid^{2}$, $\mid V_{eN}V_{\mu N}\mid$ given in eqn.(\ref{analytical4}), for possible upper limits on
the $\mathscr{B}^{exp}$ $\left(B_{c}^{-}\rightarrow \overline{B}_{s}^{0}\ell_{1}^{-}\ell_{2}^{-}\pi^{+}\right)$, of $10^{-4}$ and  $10^{-5}$.
Rather loose constraints are obtained if no assumptions regarding the mixing elements $\mid V_{\ell N}\mid^{2}, \ell=e, \mu, \tau$ are made and if the maximum values of these mixing elements 
 permitted by unitarity and global fits to oscillation data (obtained in
 eqn.(\ref{oscillation bound})) are used in $\Gamma_{N}$ evaluation. This results in the upper limits  
on the mixing elements displayed in the right panels of the Figs. \ref{1}, \ref{2} and \ref{3}, again if upper limits on the
$\mathscr{B}^{exp}$ $\left(B_{c}^{-}\rightarrow \overline{B}_{s}^{0}\ell_{1}^{-}\ell_{2}^{-}\pi^{+}\right)$ of $10^{-4}$, $10^{-5}$ are experimentally attained. 
\begin{figure}
\subfigure[]{\includegraphics[width=8cm]{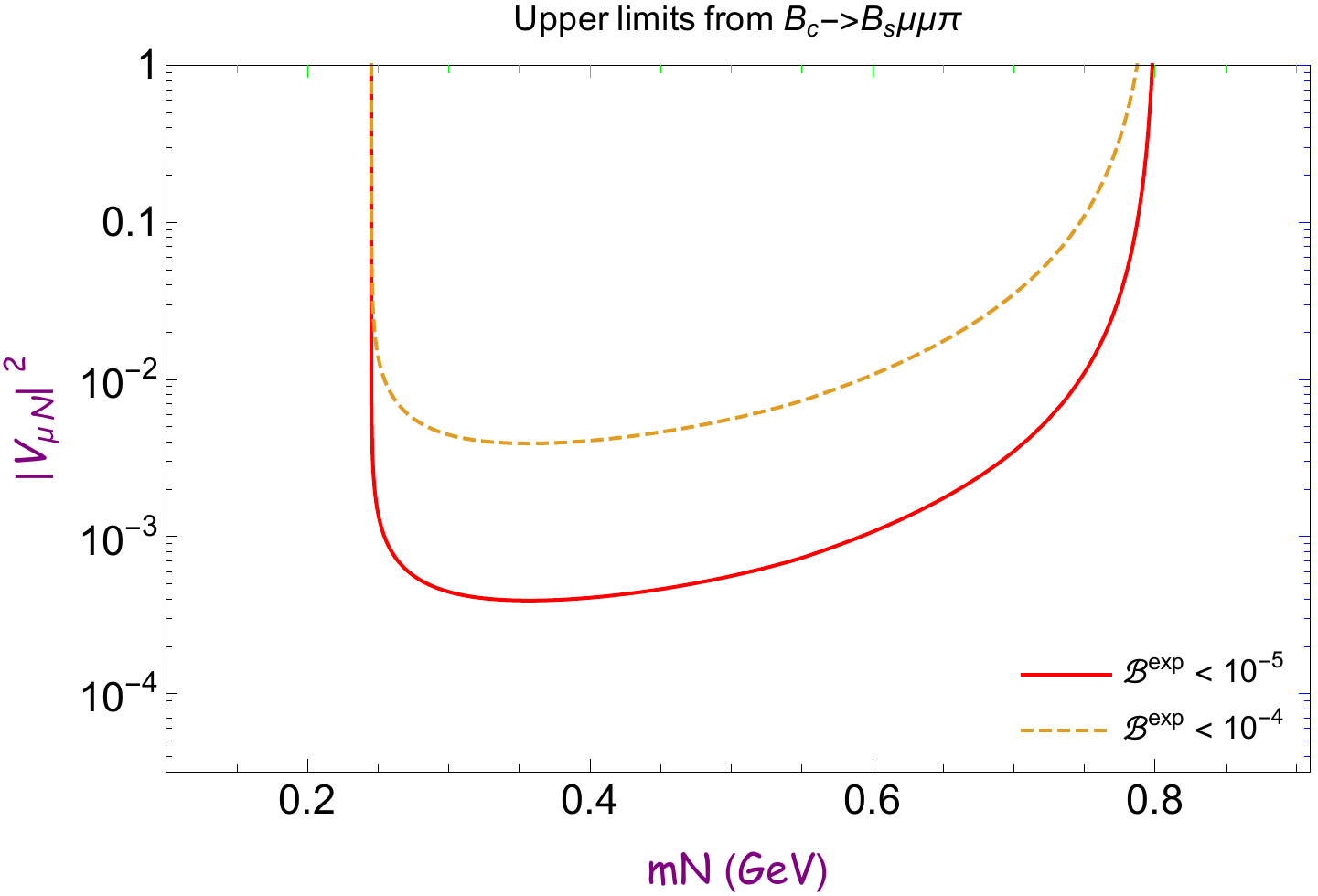}}
\subfigure[]{\includegraphics[width=8cm]{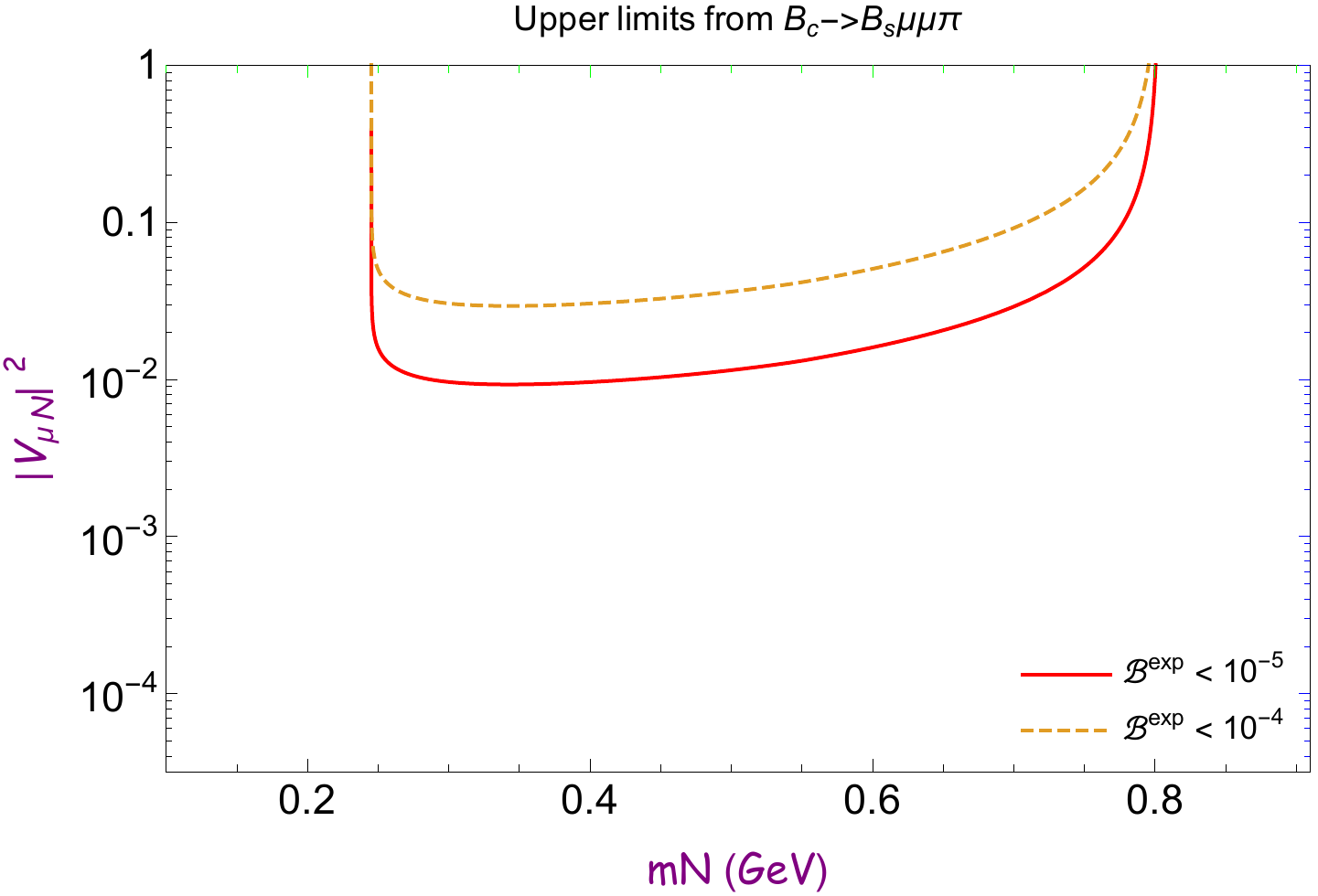}}
\caption{\small{Exclusion curves for the mixing element $\mid V_{\mu N}\mid^{2}$ from the expected upper limits for the branching fraction of the decay mode
$B_{c}^{-}\rightarrow \overline{B}_{s}^{0}\mu^{-}\mu^{-}\pi^{+}$. Both the left and right plots use the same assumption/constraints for the magnitude of the mixing elements as those in Fig.\ref{1}}}
\label{2}
\end{figure}

For the lepton flavour violating case, $\ell_{1}\neq\ell_{2}$, the mass difference of $e$ and $\mu$ results in a slight difference in the mass range allowed for $N$ (for its resonant production) for the two cases: when the electron
is produced along with the Majorana neutrino N, while muon arises from the decay of N, or vice versa. Hence, in Fig.\ref{3} we present the exclusion curves for these two cases separately. If the separation of the vertices is not easily feasible, one can just add the results of the two cases in the overlapping kinematic range.
Using 
the upper limit on the branching fractions, $\mathscr{B}^{exp}$ $\left(B_{c}^{-}\rightarrow \overline{B}_{s}^{0}\ell_{1}^{-}\ell_{2}^{-}\pi^{+}\right)$ 
$\sim 10^{-5}$, the bounds on the mixing angles obtained for $\sim (0.1<m_{N}<0.9)\,\text{GeV}$, are slightly tighter 
than those from other heavy meson decays considered in \cite{tao han,GammaN calculation}. Only the constraints from $K$ meson visible 3-body decays are tighter, but for the mass range of $\sim 0.35<m_{N}<0.90\,\text{GeV}$, our exclusion limits are either tighter or compatible with the earlier constraints. A comparison of our exclusion plots against that shown in a recent analysis on global constraints on a heavy neutrino~\cite{Global Constraints on a Heavy Neutrino}, again shows that these bounds 
could provide very tight constraints in a small range of $m_N$, beyond that excluded only by peak searches in $K$ meson decays, which is otherwise so far unconstrained.~\footnote{We wish to point out that our constraints cannot be directly compared with that in Ref.~\cite{Global Constraints on a Heavy Neutrino}, as their conservative constraints are independent of the heavy neutrino decay products.}

The reasons for this improved sensitivity are that 
the meson decay modes considered in the literature so far have been mostly 3-body decay modes involving the
annihilation of the initial meson and the 
weak annihilation vertex of all heavy mesons (except $D_s$) suffers from Cabibbo suppression. This reduces the coefficient of the
mixing elements in the decay rates, resulting in looser constraints. Hence, in spite of the mild phase space suppression this 4-body mode can result in improved exclusion limits
for the mixing angles of the heavy Majorana neutrino with the light flavour neutrinos. With a larger sample of $B_c$ events, possible at future high energy colliders, much stronger upper limits on the branching ratios would be possible, which would result in more stringent constraints on the mixing elements.

\section{Other $B_c$ Decay Modes}
\label{IV}
Although the modes $B_{c}^{-}\rightarrow \overline{B}_{s}^{0}\ell_{1}^{-}\ell_{2}^{-}\pi^{+}$, are expected to have a larger branching ratios due to the Cabibbo enhancement, however, as pointed out in the last section,
the reconstruction of the $\overline{B}_{s}^{0}$ results in a penalty of $\sim \mathcal{O}(10^{-4})$, implying  that with the limited number $B_c$ events at LHCb even in the $13/14\,\text{TeV}$ run, upper limits on the branching ratios for these modes, smaller than $10^{-5}$ may not be feasible. In fact, for the modes $B_{c}^{-}\rightarrow J/\psi\ell_{1}^{-}\ell_{2}^{-}\pi^{+}$ which are Cabibbo suppressed, but where the reconstruction of $J/\psi$ only results in a suppression factor of 
$\sim \mathcal{O}(10^{-2})$, tighter upper limits on the branching fraction $\sim \mathcal{O}(10^{-7})$ may be achievable, provided the final leptons are electrons or muons. If one of the final leptons is a tau, the upper limit may be less tighter $\sim \mathcal{O}(10^{-6})$. Also, while LHCb has already 
searched for Majorana neutrinos via the mode $B^-\rightarrow\pi^+\mu^-\mu^-$, perhaps a search through the mode $B_c^-\rightarrow\pi^+\mu^-\mu^-$ may provide tighter constraints on the mixing angles.

\subsection{$B_{c}^{-}\rightarrow J/\psi\ell_{1}^{-}\ell_{2}^{-}\pi^{+}$}
\begin{figure}
\subfigure[]{\includegraphics[width=8cm]{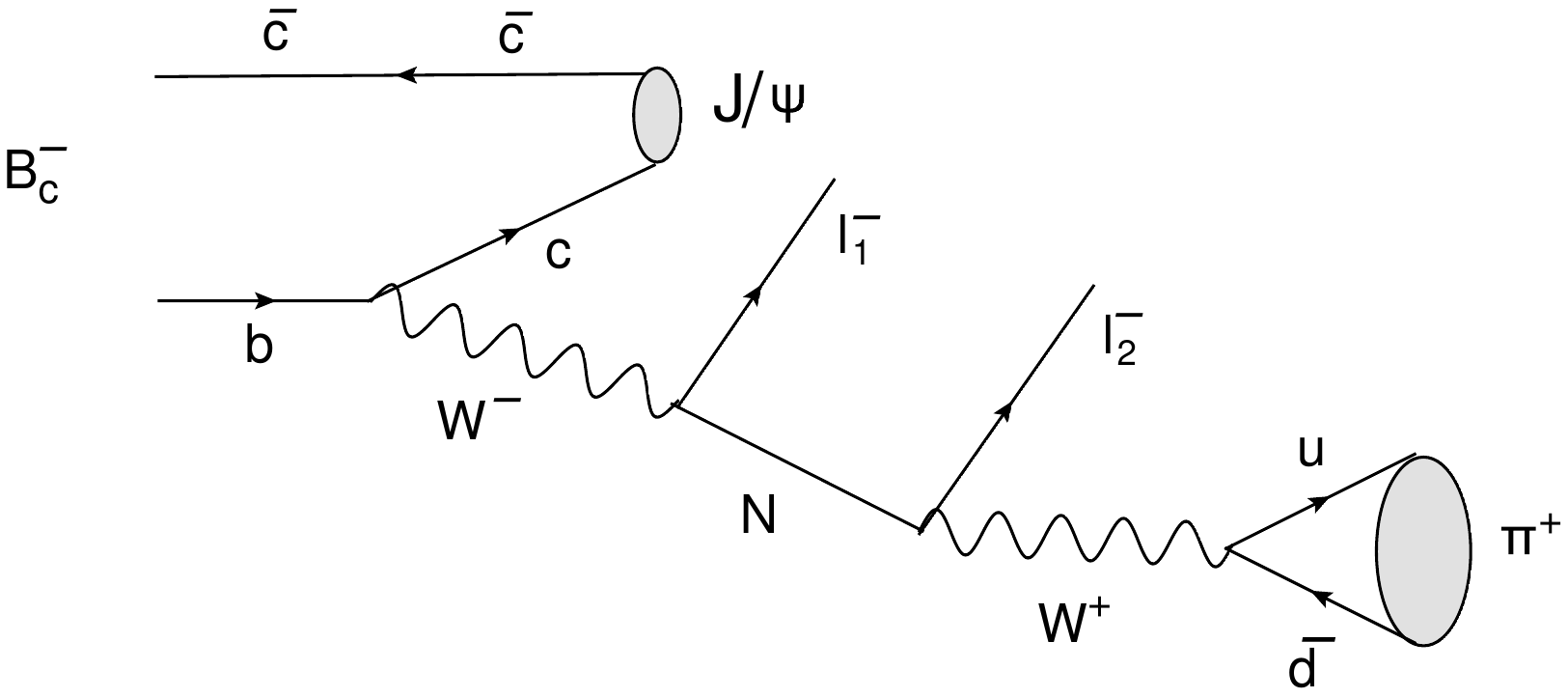}}
\subfigure[]{\includegraphics[width=8cm]{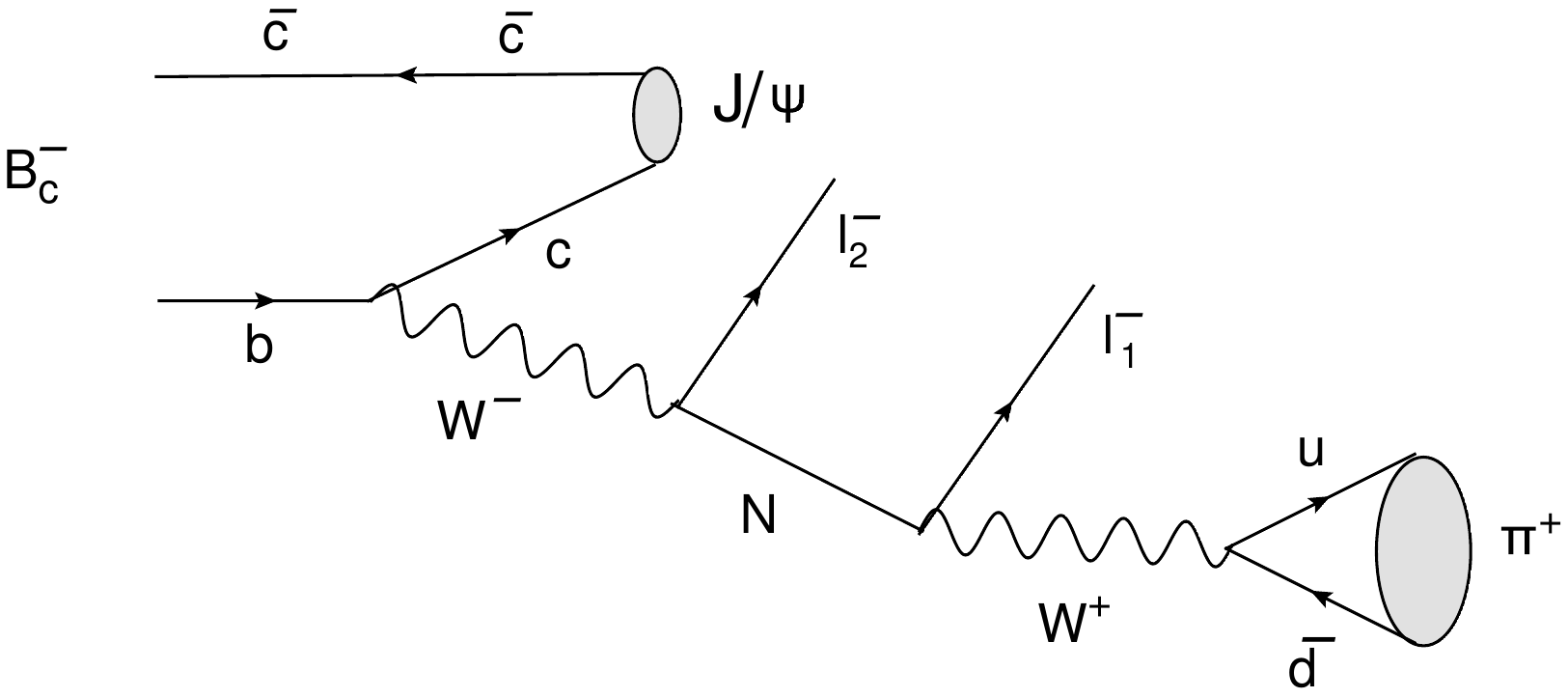}}
\caption{\small{Feynman diagrams for the decay $B_{c}^{-}\rightarrow J/\psi\ell_{1}^{-}\ell_{2}^{-}\pi^{+}$}.}
\label{feynmandiagramforBctopsi}
\end{figure}

The diagrams contributing to this decay mode are shown in Fig.~\ref{feynmandiagramforBctopsi}.
The leptonic tensor in the amplitude will have the same form as that in eqn.(\ref{leptonic}), while the hadronic tensor can be written as:
\begin{equation}
\left(\mathcal{M}_{had}\right)^{\beta\mu}=\frac{G_{F}}{\sqrt{2}}V_{cb}V_{ud}\braket{J/\psi\left(k_{1}\right)|\bar{b}\gamma^{\mu}(1-\gamma_5)c|B_{c}^{-}\left(p\right)}
\braket{\pi^{+}\left(k_{4}\right)|\bar{u}\gamma^{\beta}d|0}~,
\end{equation}
Here, the hadronic matrix element of the weak current in the $B_{c}^{-}\rightarrow J/\psi$ transition in terms of the vector and axial-vector form factors is given by, 
\begin{eqnarray}
\braket{J/\psi\left(k_{1}\right)|\mathcal{J}^{\mu}|B_{c}^{-}\left(p\right)}=\left(-F_V\epsilon^{\mu\nu\alpha\beta}\epsilon^*_\nu Q_\alpha q_\beta +i F_0^A\epsilon^{*\mu}+i F_+^A(\epsilon^*. p)Q^\mu+i F_-^A(\epsilon^*. p)q^{\mu}\right)~,
\end{eqnarray}
where, $Q=p+k_{1}$, $q=p-k_{1}$, and $\epsilon$ is the polarization vector of the $J/\psi$ meson.
The form factors, $F_V, F_0^A, F_+^A\,\text{and}\, F_-^A$ have been estimated using QCD sum rules in Ref.~\cite{form factors}, with the values from zero recoil evolved with the pole dependence:  
\begin{equation}
F_{i}\left(q^{2}\right)=\frac{F_{i}(0)}{1-\frac{q^{2}}{M_{i,pole}^{2}}}~,
\end{equation} 
with the numerical values: $F_V(0)=0.11\,\text{GeV}^{-1}, F_0^A=5.9\,\text{GeV}, F_+^A=-0.074\,\text{GeV}^{-1}\,\text{and}\,F_-^A=0.12 \,\text{GeV}^{-1}$; while the pole mass used in each of the vector/axial-vector form
factors for $B_c\rightarrow \overline{c}c$ is $4.5\,\text{GeV}$.
We evaluate the four-body decay rate for this mode using the procedure analogous to that followed for the $B_{c}^{-}\rightarrow \overline{B}_{s}^{0}\ell_{1}^{-}\ell_{2}^{-}\pi^{+}$ decay mode, i.e.,
using the narrow width approximation for $N$ and the phase space given in eqn.(\ref{phasespace}). Of course, due to the presence of larger number of form factors, the matrix element mod-squared appears more complicated.   
The bounds on the mixing elements are also derived in a similar fashion, using constraints similar to that given in eqn.(\ref{analytical4}), with the corresponding parameters appropriately defined in terms of the theoretical branching fractions and the experimental upper limits for the $B_{c}^{-}\rightarrow J/\psi\ell_{1}^{-}\ell_{2}^{-}\pi^{+}$ mode.  Note however, that the mass difference between that of $B_c$ and $J/\psi$ will allow neutrino masses up to over $3\,\text{GeV}$ to be on shell.
This not only allows us to constrain $\mid V_{eN}\mid^2,\mid V_{\mu N}\mid^2$ and $\mid V_{eN}V_{\mu N}\mid$ over a bigger mass range, but  exclusion curves for  $\mid V_{eN}V_{\tau N}\mid, \mid V_{\mu N}V_{\tau N}\mid$ can also be provided for heavy neutrino masses beyond the region probed via tau decays.

\subsection{$B_{c}^{-}\rightarrow \pi^{+}\ell_{1}^{-}\ell_{2}^{-}$}
\begin{figure}
	\subfigure[]{\includegraphics[width=8cm]{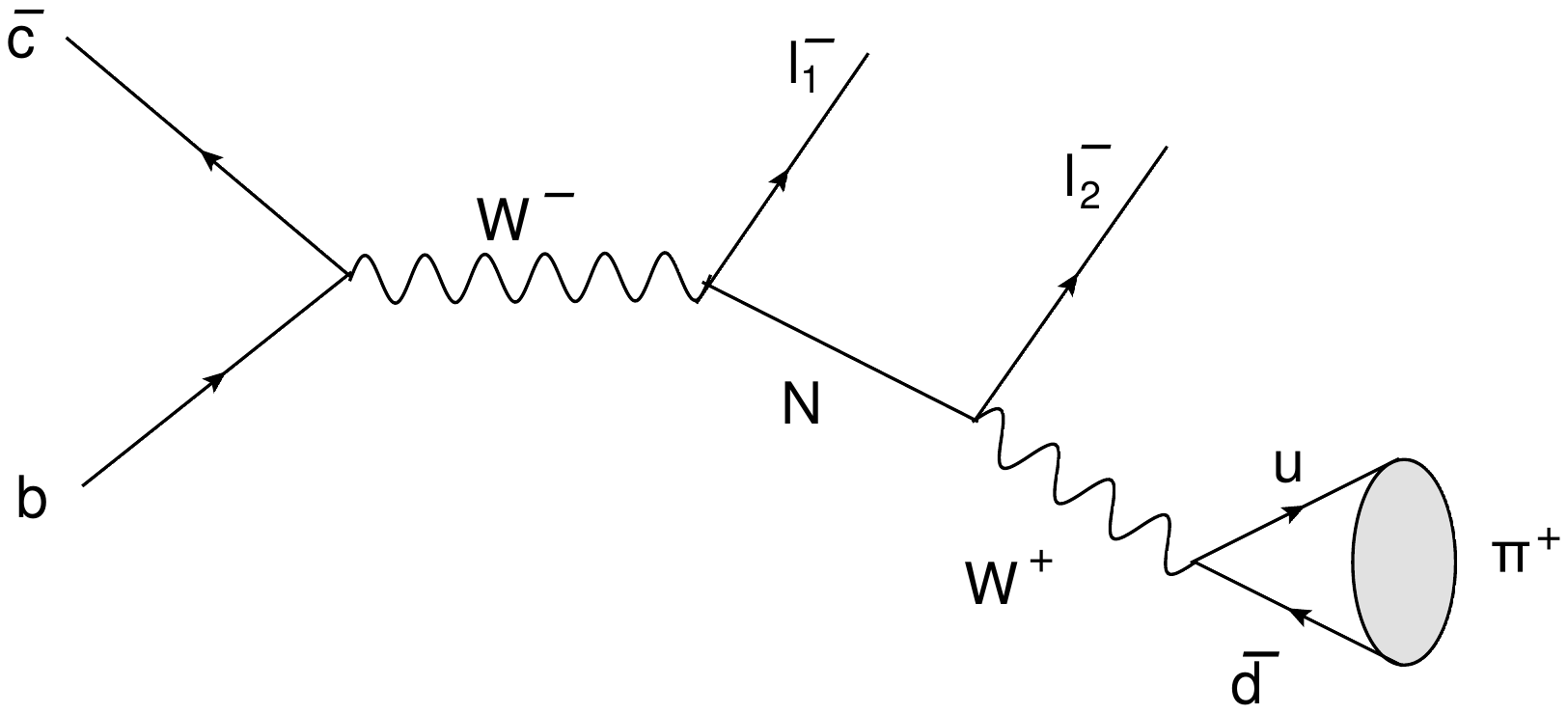}}
	\subfigure[]{\includegraphics[width=5cm]{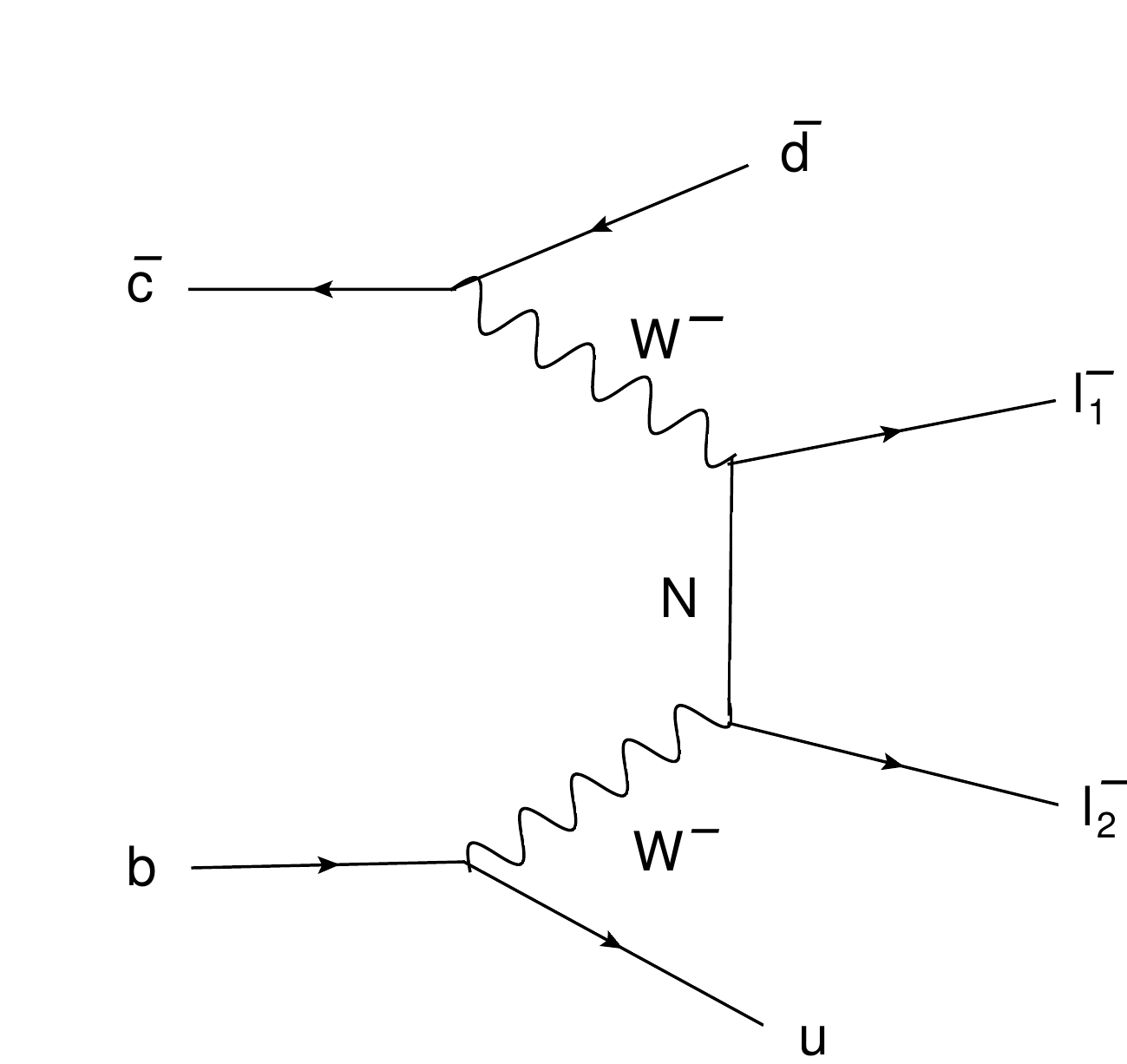}}
	\caption{\small{Feynman diagrams for the decay $B_{c}^{-}\rightarrow\pi^{+}\ell_{1}^{-}\ell_{2}^{-}$}.}
	\label{diagramforBctopill}
\end{figure}
While the number of $B_c$ events at LHCb are expected to be smaller than the number of $B^\pm$ events, still this mode being less  suppressed with respect to $B^{-}\rightarrow \pi^{+}\ell_{1}^{-}\ell_{2}^{-}$,
could possibly result in tighter constraints on the mixing angles. The diagrams contributing to this process are shown in Figs.~\ref{diagramforBctopill}. Apart from the s-channel diagram (a), there is also a t-channel diagram, where the off-shell heavy neutrino contributes. However, since this diagram is highly suppressed due to CKM suppression, as well as due to absence of resonant enhancement, we only include the dominant contribution of Fig.~\ref{diagramforBctopill}(a) (including that for the two leptons 
exchanged). The large mass difference between that of $B_c$ and $\pi$ meson allows both final leptons to be taus also. With only pion and electrons/muons as the final state particles, this mode should be easy to reconstruct, however, for the case of one or both of
the leptons being a tau, the reconstruction will involve accounting for the tau branching fraction to the final state through which it is seen. 
The even wider range allowed for the heavy neutrino mass, also allows upper limit on $\mid V_{\tau N}\mid^2$, which is unconstrained by any of the $\tau$ or other meson decays.
 
\begin{figure}
\subfigure[]{\includegraphics[width=8cm]{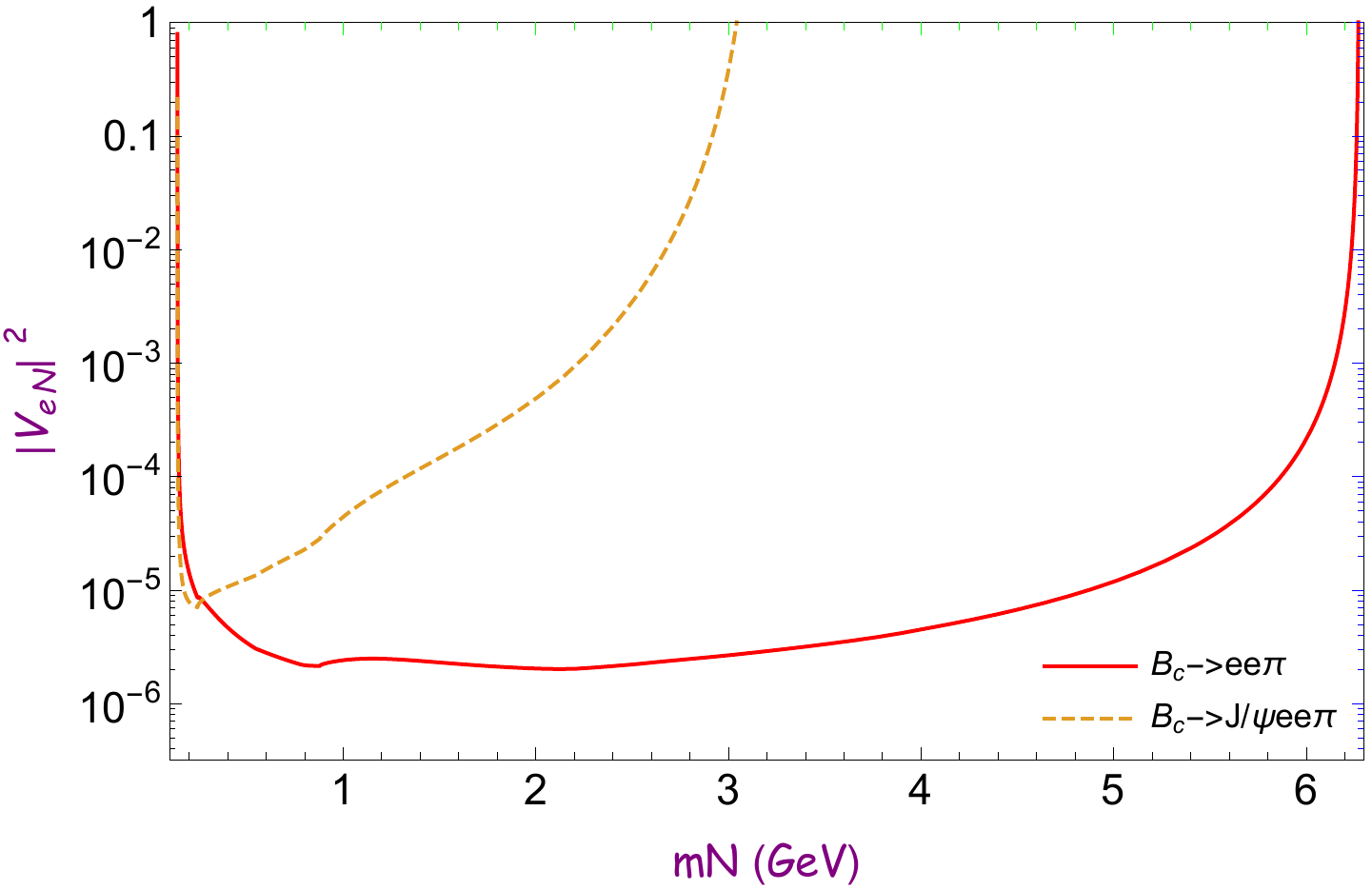}}
\subfigure[]{\includegraphics[width=8cm]{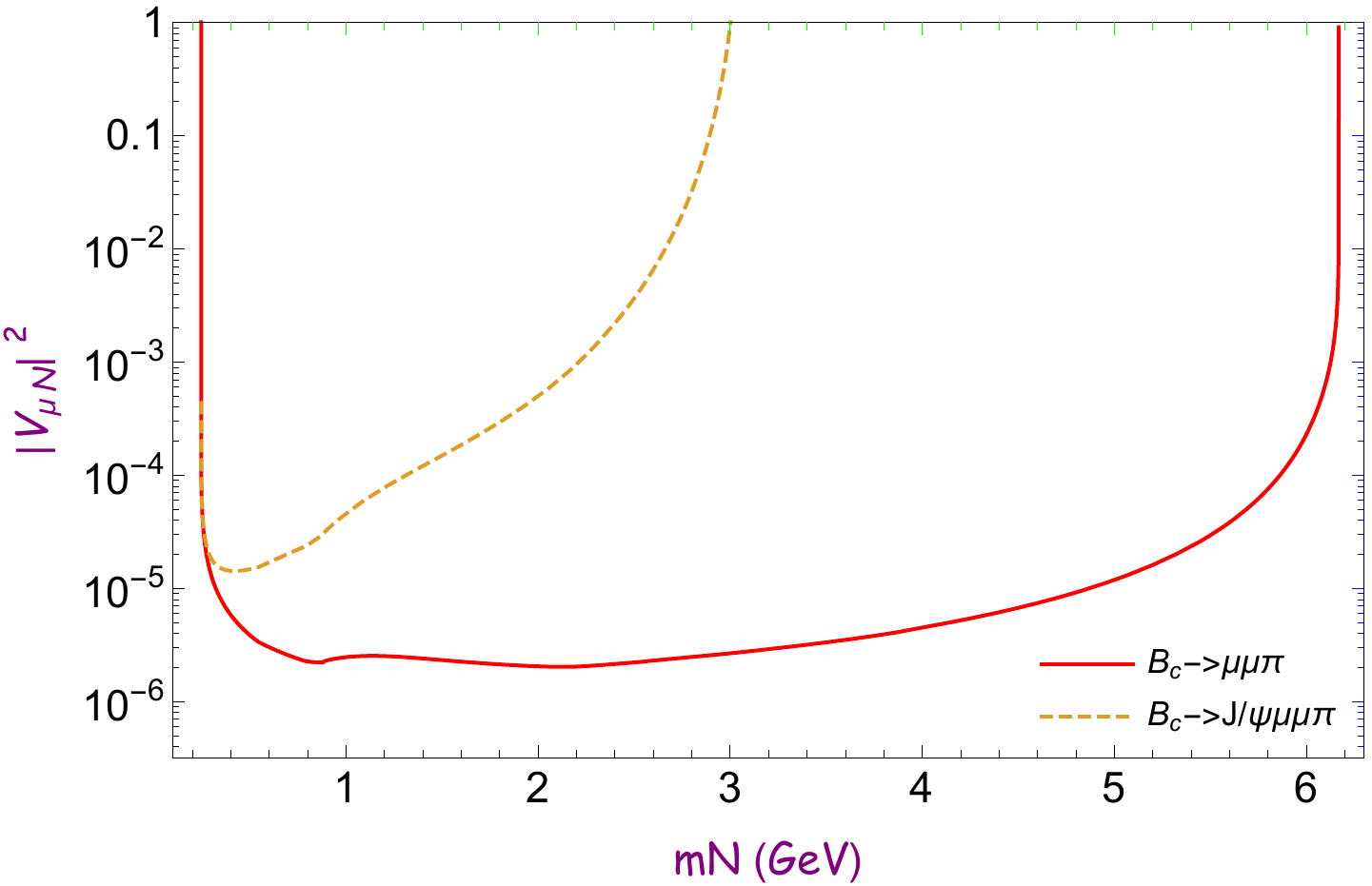}}
\subfigure[]{\includegraphics[width=8cm]{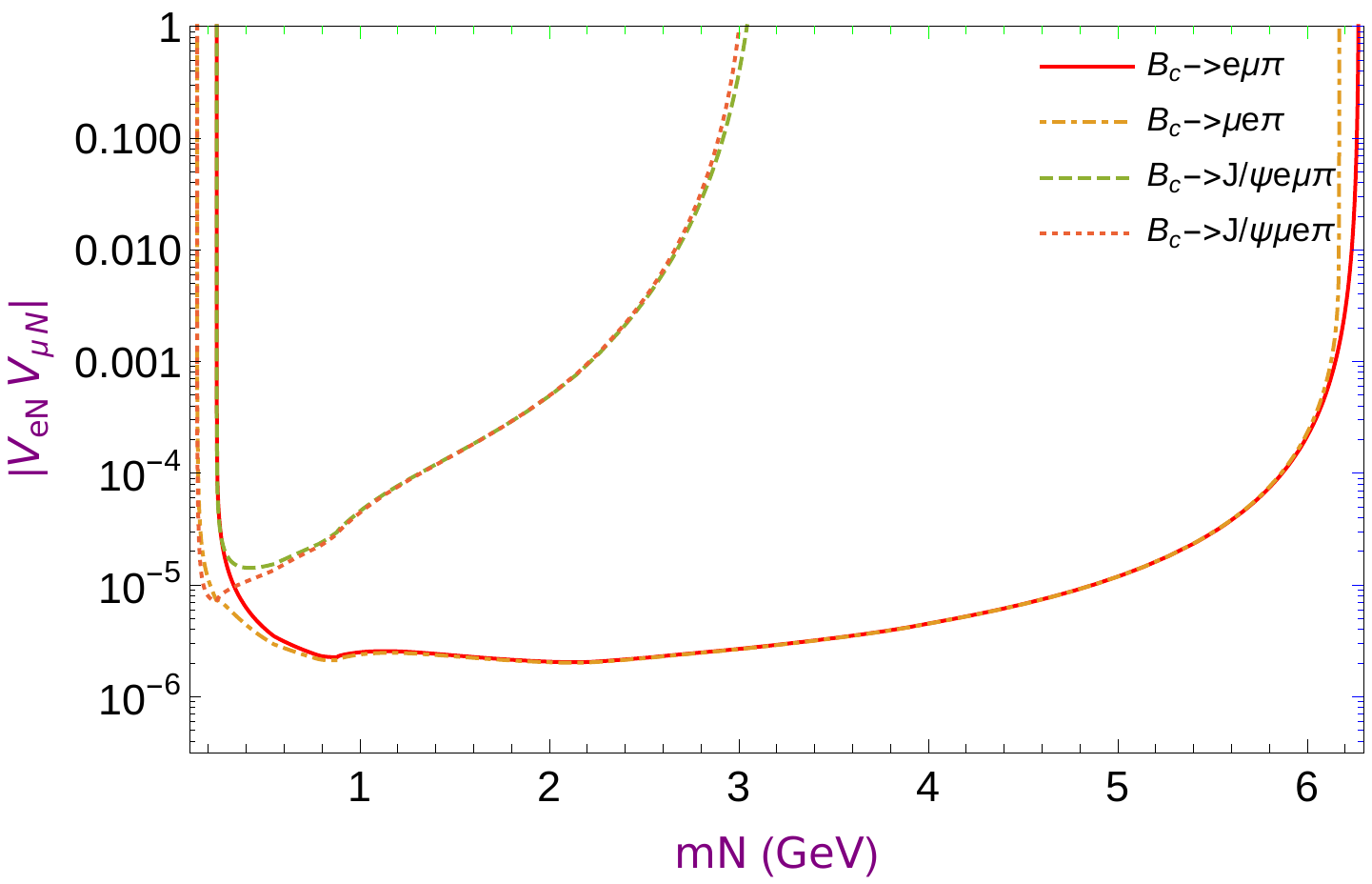}}
\caption{\small{Exclusion curves for the mixing element $\mid V_{\ell_1 N} V_{\ell_2 N}\mid$ from upper limits for the branching fraction
$\mathscr{B}\left(B_{c}^{-}\rightarrow J/\psi\ell_1^{-}\ell_2^{-}\pi^{+}\right)\sim10^{-7}$ and $\mathscr{B}\left(B_{c}^{-}\rightarrow\pi^{+}\ell_{1}^{-}\ell_{2}^{-}\right)\sim 10^{-9}$. Notation regarding the ordering of the leptons is the same as that described in Sec.~\ref{III}}}
\label{VeNVmuNVeNVmuN}
\end{figure}

\begin{figure}
\subfigure[]{\includegraphics[width=8cm]{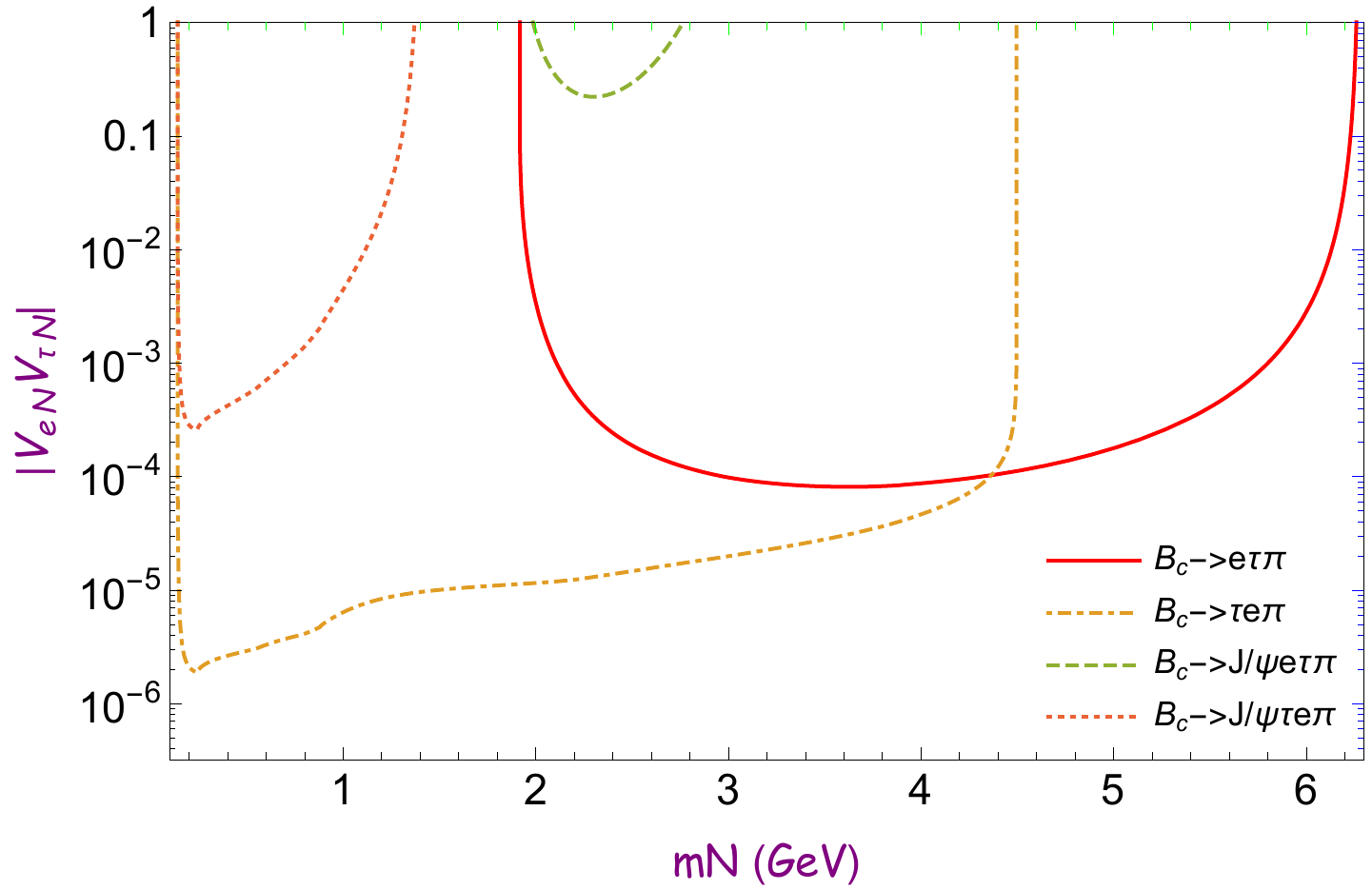}}
\subfigure[]{\includegraphics[width=8cm]{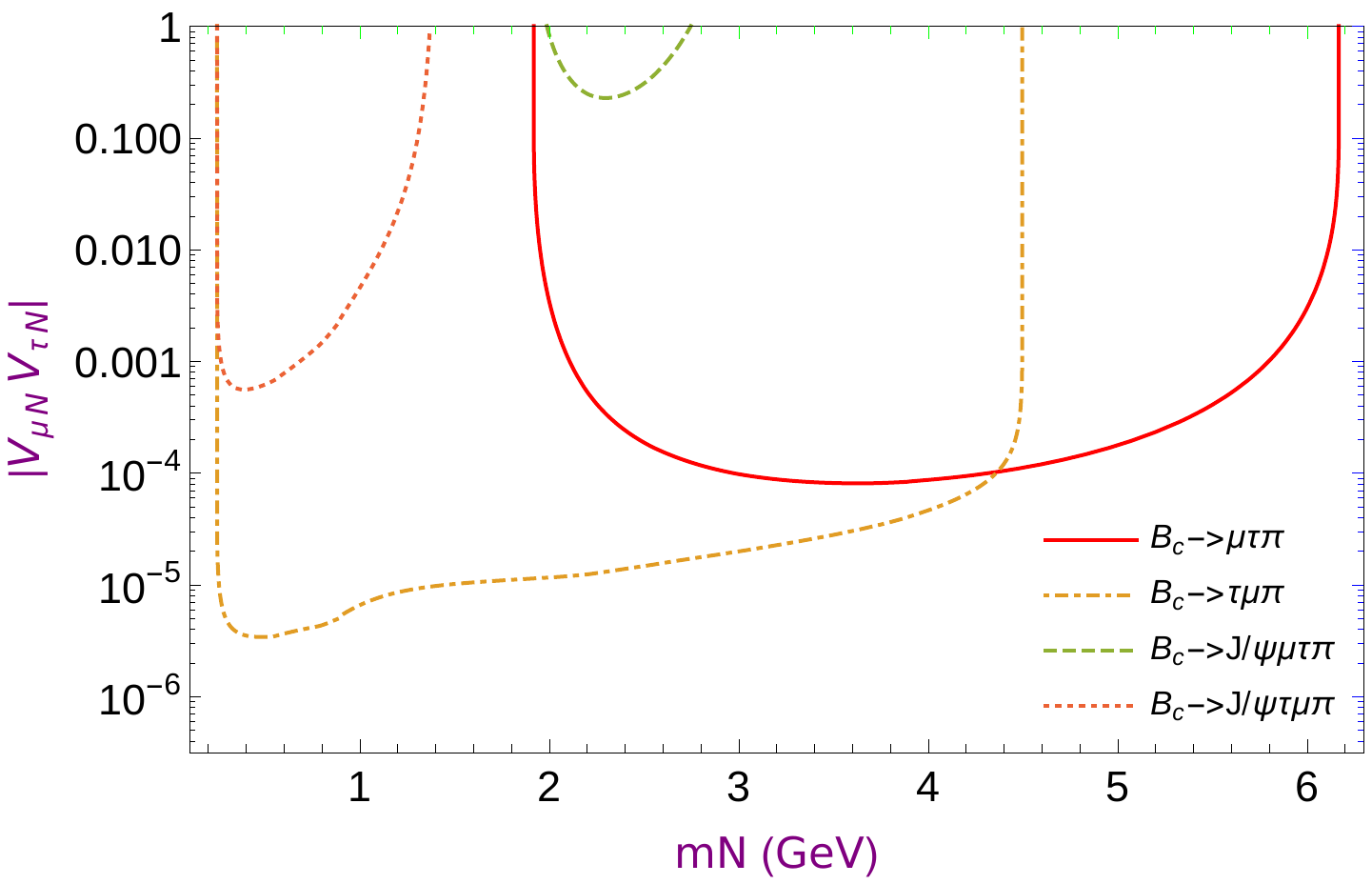}}
\subfigure[]{\includegraphics[width=8cm]{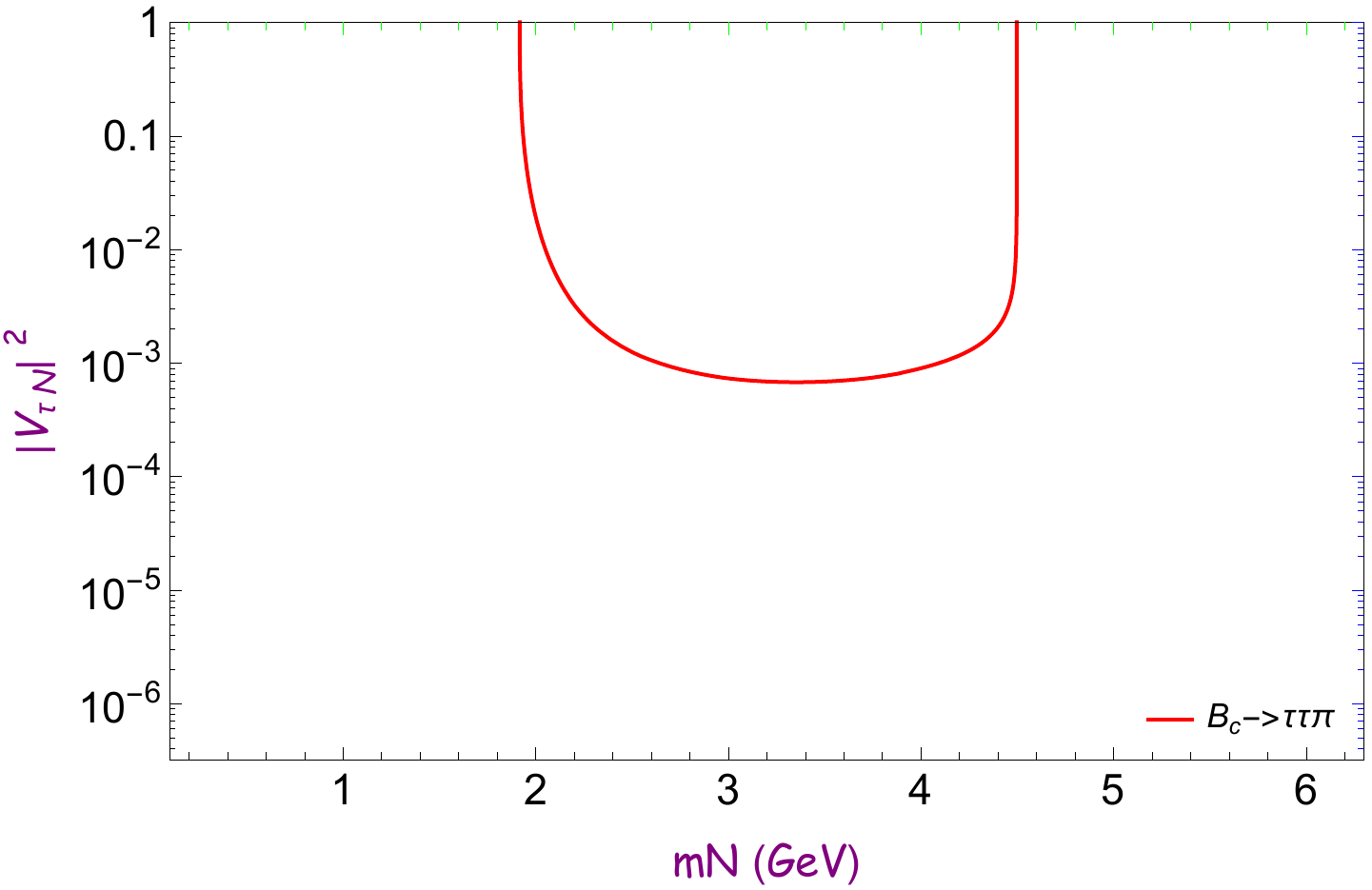}}
\caption{\small{Exclusion curves for the mixing element $\mid V_{\ell_1 N} V_{\ell_2 N}\mid$.
For (a), one of the leptons is an electron while the second one is a tau; the upper limits used are: $\mathscr{B}\left(B_{c}^{-}\rightarrow J/\psi\ell_1^{-}\ell_2^{-}\pi^{+}\right)\sim 10^{-6}$, 
$\mathscr{B}\left(B_{c}^{-}\rightarrow\pi^{+}\ell_{1}^{-}\ell_{2}^{-}\right)\sim 10^{-8}$,
(b) corresponds to the case of one muon and one tau, again using the upper limits: $\mathscr{B}\left(B_{c}^{-}\rightarrow J/\psi\ell_1^{-}\ell_2^{-}\pi^{+}\right)\sim 10^{-6}$,
$\mathscr{B}\left(B_{c}^{-}\rightarrow\pi^{+}\ell_{1}^{-}\ell_{2}^{-}\right)\sim 10^{-8}$ and for (c), both final leptons are taus and the expected upper limit
for $\mathscr{B}\left(B_{c}^{-}\rightarrow\pi^{+}\ell_{1}^{-}\ell_{2}^{-}\right)\sim 10^{-7}$.}}
\label{VtauNVeNVtauNVmuNVtauN}
\end{figure}

In Fig.~\ref{VeNVmuNVeNVmuN} (a), (b) and (c) we show the exclusion curves for $\mid V_{eN}\mid^2$, $\mid V_{\mu N}\mid^2$ and $\mid V_{eN}V_{\mu N}\mid$ respectively, obtained from the expected upper limits
of $\mathscr{B}\left(B_{c}^{-}\rightarrow J/\psi\ell_{1}^{-}\ell_{2}^{-}\pi^{+}\right)\sim 10^{-7}$  and $\mathscr{B}\left(B_{c}^{-}\rightarrow\pi^{+}\ell_{1}^{-}\ell_{2}^{-}\right)\sim 10^{-9}$
($\ell_{1},\ell_{2}=e\,\,\text{or}\,\,\mu$), at LHCb with $\sim 10^{10}$ $B_c$ events.
If one or both of the leptons is a tau, then it's reconstruction would lead to looser upper limits on the branching fraction.  Fig.~\ref{VtauNVeNVtauNVmuNVtauN}(a) shows the exclusion curves for $\mid V_{eN}V_{\tau N}\mid$,
while that for $\mid V_{\mu N}V_{\tau N}\mid$ are displayed in Fig.~\ref{VtauNVeNVtauNVmuNVtauN}(b), corresponding to the upper limits: $\mathscr{B}\left(B_{c}^{-}\rightarrow J/\psi\ell_1^{-}\ell_2^{-}\pi^{+}\right)\sim 10^{-6}$ and
$\mathscr{B}\left(B_{c}^{-}\rightarrow\ell_{1}^{-}\ell_{2}^{-}\pi^{+}\right)\sim 10^{-8}$ when, $\ell_{1}^{-}$ or $\ell_{2}^{-}$ is a $\tau^-$. Fig.~\ref{VtauNVeNVtauNVmuNVtauN}(c) shows the exclusion curve
for $\mid V_{\tau N}\mid^2$ corresponding to an upper limit of $\mathscr{B}\left(B_{c}^{-}\rightarrow\pi^{+}\tau^{-}\tau^-\right)\sim 10^{-7}$         
 
Note that $\mid V_{\tau N}\mid^2$ is very loosely constrained, with some limits from CHARM~\cite{CHARM1,CHARM2}, NOMAD~\cite{NOMAD} and DELPHI~\cite{DELPHI} collaborations, but with the mass range of $\sim(0.3-5.0)\,\text{GeV}$ almost unconstrained. The $B_{c}^{-}\rightarrow\pi^{+}\tau^-\tau^-$ mode partially fills up this gap in providing exclusion limits in part of this range.
    
In each of the above studies the Majorana sterile neutrino produced in the $B_{c}$ decay is assumed to propagate as a real particle and then decay after a certain distance
from the production point. In the exclusion limits obtained on the mixing elements above, we assumed an idealized detector, where this distance
lies within the detector length and hence the probability of this production and decay of the heavy neutrino within the detector is unity. In practice one may
need to introduce a more realistic probability factor, which could possibly weaken the constraints on the mixing elements. Estimation of this effect will depend on
the specific experimental set up, the momenta carried by the heavy neutrino which would depend on that of the decaying $B_c$ meson etc. Hence this can be properly
incorporated only by the respective experimental collaborations in their data analysis. In fact, LHCb has indeed accounted for this in their analysis of a few LNV
$B$ decay modes, for a Majorana neutrino of mass of $2-3\,\text{GeV}$~\cite{LHCb2012}.

\section{Conclusions}
\label{V}
We propose several $B_c$ decay modes for Majorana neutrino searches. The $B_{c}$ meson is unique in being the only meson with two heavy quarks of different flavour, allowing weak decays not only of the $b$
quark but also the $c$ quark. The $b$ quark decays are always Cabibbo suppressed, with $\lambda^2$ or $\lambda^3$ suppression for $b\to c$ or $b\to u$ transitions
respectively. The charm quark decay on the other hand can be Cabibbo favoured. Hence the amplitude for $B_{c}^{-}\rightarrow \overline{B}_{s}^{0}\ell_{1}^{-}\ell_{2}^{-}\pi^{+}$, ($\ell_{1}, \ell_{2}=e, \mu$) decays can be enhanced. These four-body decay modes
involve transition form factors rather than decay constants that appear in case of annihilation of the decaying meson, as is the case for the 3-body
meson decays extensively studied for Majorana neutrino searches in the literature. To avoid model dependence and theoretical uncertainties, we suggest that these form factors
be measured using the semileptonic mode,  $B_{c}^{-}\rightarrow\overline{B}_{s}^{0}\mu^-\overline{\nu}$. For a Majorana neutrino that lies in the mass range that
allows it to be on the mass shell, there is also a resonant enhancement of the process. A search for  Majorana neutrinos via these rare modes which are expected to
have larger branching fractions, appears more feasible. Even a non-observation can result in exclusion curves for the mixing angles of the heavy Majorana singlet
with the flavour eigenstates, corresponding to possible upper limits for the branching fractions.
These constriants are mostly tighter than those obtained from other heavy meson decay modes in earlier studies and the mass range probed lies beyond the range with stringent constraints from experimental bounds on three-body Kaon LNV decays.

In spite of the Cabibbo enhancement for the $B_{c}^{-}\rightarrow\overline{B}_{s}^{0}\ell_{1}^{-}\ell_{2}^{-}\pi^{+}$ modes, the reconstruction of the $B_s$
leads one to expect less stringent upper limits for these modes compared to that for $B_{c}^{-}\rightarrow J/\psi\ell_{1}^{-}\ell_{2}^{-}\pi^{+}$ modes where the $J/\psi$ can be reconstructed more
easily via the $\mu^+\mu^-$ mode. Similarly the reconstruction of the $B_{c}^{-}\rightarrow \pi^{+}\ell_{1}^{-}\ell_{2}^{-}$ mode would be less demanding. This along with the phase space enhancement of the latter
two modes may result in much tighter (by almost an order of magnitude) exclusion curves for the mixing elements,
$\mid V_{e N}\mid^2$, $\mid V_{\mu N}\mid^2$, $\mid V_{e N} V_{\mu N}\mid$. Further, for $\mid V_{eN}V_{\tau N}\mid, \mid V_{\mu N}V_{\tau N}\mid$, on which bounds exist only from tau decays, exclusion curves for masses upto about $6\,\text{GeV}$ can be provided. Also, upper limits for $\mid V_{\tau N}\mid^2$  can be obtained in the mass range $(0.3-
5.0)\,\text{GeV}$, where it is so far unconstrained.

\acknowledgements 

N.S. thanks Vanya Belyaev for communication regarding the expected $B_c$ production cross-section at LHCb in the $13/14\,\text{TeV}$ run. The authors thank Patrick Koppenburg for his valuable inputs and suggestions and appreciate comments from Andrew Kobach, G. López Castro, V.V. Kiselev, Sandip Pakvasa and N.G. Deshpande.      
\section{Appendix}
To describe the kinematics of four-body decays, five independent variables are required. We choose the independent variables to be, $M_{12}^{2}$, $M_{34}^{2}$,
$\theta_{12}$, $\theta_{34}$ and $\phi$, which for the processes,  $B_{c}^{-}(p)\rightarrow \overline{B}_{s}^{0}(k_1)\ell_{1}^{-}(k_2)\ell_{2}^{-}(k_3)\pi^{+}(k_4)$ or  $B_{c}^{-}(p)\rightarrow J/\psi(k_1)\ell_{1}^{-}(k_2)\ell_{2}^{-}(k_3)\pi^{+}(k_4)$ are defined as:
\begin{equation}
\begin{split}
&M_{12}^{2}=\left(k_{1}+k_{2}\right)^{2}\,\,;\,\,\,M_{34}^{2}=\left(k_{3}+k_{4}\right)^{2};\\
&\text{cos}\theta_{12}=\frac{\hat{v}.\vec{k}_{1}}{\mid\vec{k}_{1}\mid}\,\,;\,\,\,\text{cos}\theta_{34}=\frac{-\hat{v}.\vec{k}_{3}}{\mid\vec{k}_{3}\mid},
\end{split}
\end{equation}
$\phi$ is the angle between the normals to the planes defined in the $B_{c}$ rest frame by the $\overline{B}_{s}^{0}(J/\psi)\ell_{1}$ pair and the $\ell_{2}\pi^{+}$ pair. 
The ranges of the angular variables are $0\leq\theta_{12}\leq\pi$, $0\leq\theta_{34}\leq\pi$, and $-\pi\leq\phi\leq\pi$.
\begin{figure}
	\label{kinematic planes}
	\centering
	\includegraphics[width=0.7\textwidth]{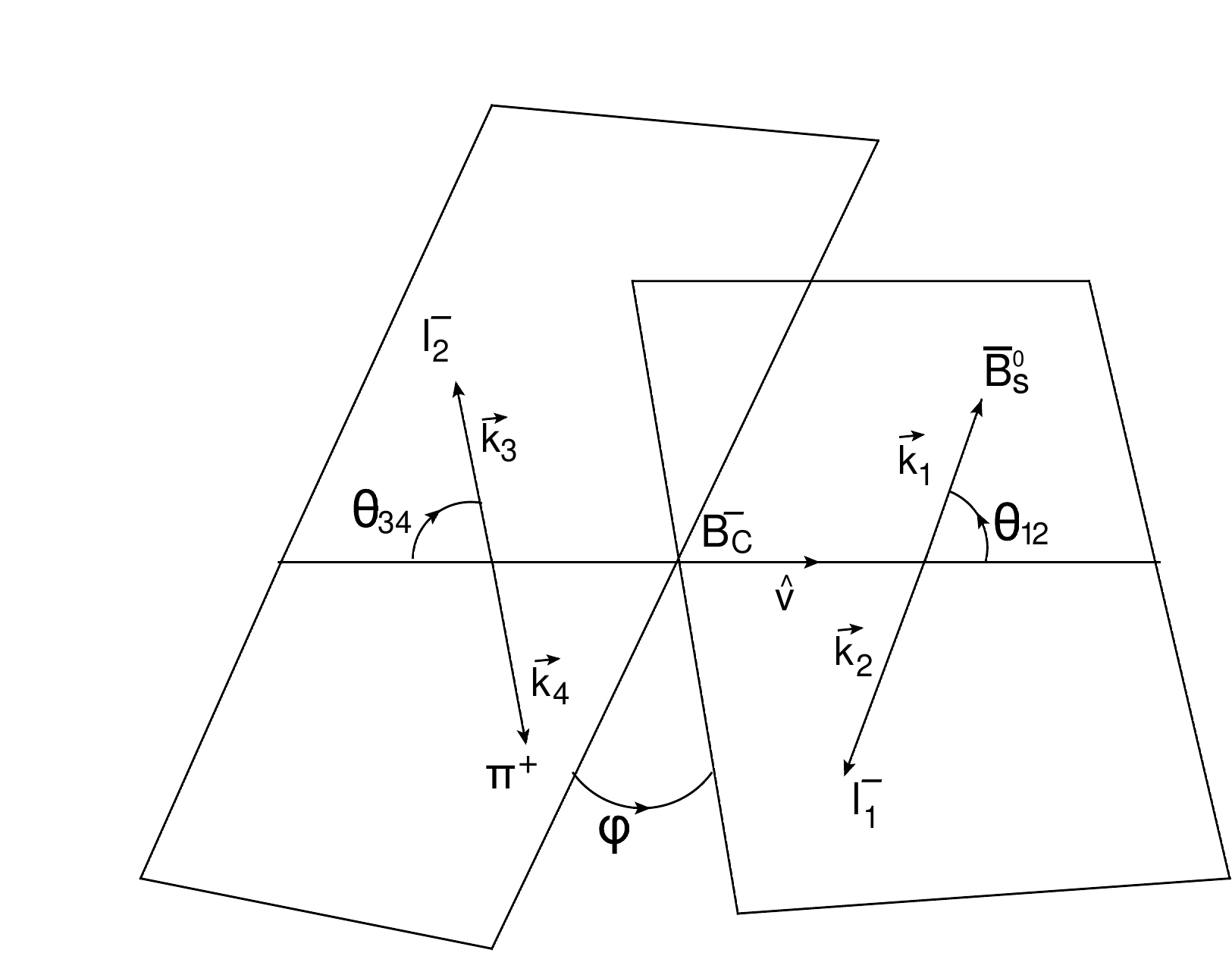}
	\caption{\small{Kinematics of four-body decays $B_{c}^{-}\rightarrow \overline{B}_{s}^{0}\ell_{1}^{-}\ell_{2}^{-}\pi^{+}$ in the $B_{c}$ rest frame}.}
\end{figure}

To evaluate the decay rate for the 4-body LNV $B_{c}^{-}\rightarrow \overline{B}_{s}^{0}\ell_{1}^{-}\ell_{2}^{-}\pi^{+}$ mode,
the mod squared of the matrix element specified in eqn. (\ref{amplitude1}) is expressed in terms of the dot products of the momenta of the final state particles as:
\begin{eqnarray}
&\sum \mid \mathcal{M}\mid^{2}=
G_{F}^{4}m_{N}^{2}\mid V_{cs}\mid^{2}\mid V_{ud}\mid^{2}\mid V_{\ell_{1}N}V_{\ell_{2}N}\mid^{2}f_{\pi}^{2}\frac{\pi}{m_{N}\Gamma_{N}}\delta\left(p_{N}^{2}-m_{N}^{2}\right)\nonumber\\
&(8\left(F_{+}^{2}+2F_{+}F_{-}+F_{-}^{2}\right)
(m_{4}^{2}m^{2}\left(k_{2}.k_{3}\right)-2m^{2}\left(k_{2}.k_{4}\right)\left(k_{3}.k_{4}\right)+4\left(k_{2}.p\right)\left(k_{3}.k_{4}\right)\left(k_{4}.p\right)\nonumber\\
&-2m_{4}^{2}\left(k_{2}.p\right)\left(k_{3}.p\right))+8\left(F_{+}^{2}-2F_{+}F_{-}+F_{-}^{2}\right)(m_{4}^{2}m_{1}^{2}\left(k_{2}.k_{3}\right)-2m_{1}^{2}\left(k_{2}.k_{4}\right)\left(k_{3}.k_{4}\right)\nonumber\\
&+4\left(k_{1}.k_{2}\right)\left(k_{3}.k_{4}\right)\left(k_{4}.k_{1}\right)
-2m_{4}^{2}\left(k_{1}.k_{2}\right)\left(k_{1}.k_{3}\right))+16\left(F_{+}^{2}-F_{-}^{2}\right)(m_{4}^{2}\left(k_{2}.k_{3}\right)\left(p.k_{1}\right)\nonumber\\
&-2\left(k_{2}.k_{4}\right)\left(k_{3}.k_{4}\right)\left(p.k_{1}\right)+2\left(p.k_{2}\right)\left(k_{3}.k_{4}\right)\left(k_{1}.k_{4}\right)
-m_{4}^{2}\left(p.k_{2}\right)\left(k_{1}.k_{3}\right)+2\left(k_{1}.k_{2}\right)\left(k_{3}.k_{4}\right)\nonumber\\
&\left(p.k_{4}\right)-m_{4}^{2}\left(k_{1}.k_{2}\right)\left(p.k_{3}\right)))+\left(k_{2}\leftrightarrow k_{3},\,m_{2}\leftrightarrow m_{3}\right)~.
\end{eqnarray}
Following are the explicit form of the four momenta  of the final state particles $\overline{B}_{s}^{0}\left(k_{1}\right)$,
$\ell_{1}^{-}\left(k_{2}\right)$, $\ell_{2}^{-}\left(k_{3}\right)$ and $\pi^{+}\left(k_{4}\right)$ in the $B_{c}$ rest frame,
\begin{equation}
\begin{split}
p&=[m,0,0,0];\\
k_{1}^{\mu}&=\Big[\frac{\sqrt{M_{12}^{2}+X^{2}}}{2M_{12}^{2}}\left(M_{12}^{2}+m_{1}^{2}-m_{2}^{2}\right)+\frac{X}{2}\cos(\theta_{12})\lambda^{\frac{1}{2}}
\left(1,\frac{m_{1}^{2}}{M_{12}^{2}},\frac{m_{2}^{2}}{M_{12}^{2}}\right),\\
&\frac{1}{2}M_{12}\lambda^{\frac{1}{2}}\left(1,\frac{m_{1}^{2}}{M_{12}^{2}},\frac{m_{2}^{2}}{M_{1}^{2}}\right)\sin(\theta_{12}),0,
\frac{1}{2}\sqrt{M_{1}^{2}+X^{2}}\cos(\theta_{12})\\
&\lambda^{\frac{1}{2}}\left(1,\frac{m_{1}^{2}}{M_{12}^{2}},\frac{m_{2}^{2}}{M_{12}^{2}}\right)+\frac{X}{2M_{12}^{2}}\left(M_{12}^{2}+m_{1}^{2}-m_{2}^{2}\right)\Big];\\
k_{2}^{\mu}&=\Big[\frac{\sqrt{M_{12}^{2}+X^{2}}}{2M_{12}^{2}}\left(M_{12}^{2}+m_{2}^{2}-m_{1}^{2}\right)-\frac{X}{2}\cos(\theta_{12})\lambda^{\frac{1}{2}}
\left(1,\frac{m_{1}^{2}}{M_{12}^{2}},\frac{m_{2}^{2}}{M_{12}^{2}}\right),\\
&-\frac{1}{2}M_{12}\lambda^{\frac{1}{2}}\left(1,\frac{m_{1}^{2}}{M_{12}^{2}},\frac{m_{2}^{2}}{M_{12}^{2}}\right)\sin(\theta_{12}),0,
-\frac{1}{2}\sqrt{M_{12}^{2}+X^{2}}\cos(\theta_{12})\\
&\lambda^{\frac{1}{2}}\left(1,\frac{m_{1}^{2}}{M_{12}^{2}},\frac{m_{2}^{2}}{M_{12}^{2}}\right)+\frac{X}{2M_{12}^{2}}\left(M_{12}^{2}+m_{2}^{2}-m_{1}^{2}\right)\Big];\\
k_{3}^{\mu}&=\Big[\frac{\sqrt{M_{34}^{2}+X^{2}}}{2M_{34}^{2}}\left(M_{34}^{2}+m_{3}^{2}-m_{4}^{2}\right)-\frac{X}{2}\cos(\theta_{34})\lambda^{\frac{1}{2}}
\left(1,\frac{m_{3}^{2}}{M_{34}^{2}},\frac{m_{4}^{2}}{M_{34}^{2}}\right),\\
&\frac{1}{2}M_{34}\lambda^{\frac{1}{2}}\left(1,\frac{m_{3}^{2}}{M_{34}^{2}},\frac{m_{4}^{2}}{M_{34}^{2}}\right)\sin(\theta_{34})\cos(\phi)
,\frac{1}{2}M_{34}\lambda^{\frac{1}{2}}\left(1,\frac{m_{3}^{2}}{M_{34}^{2}},\frac{m_{4}^{2}}{M_{34}^{2}}\right)\sin(\theta_{34})\sin(\phi),\\
&\frac{1}{2}\sqrt{M_{34}^{2}+X^{2}}\cos(\theta_{34})
\lambda^{\frac{1}{2}}\left(1,\frac{m_{3}^{2}}{M_{34}^{2}},\frac{m_{4}^{2}}{M_{34}^{2}}\right)-\frac{X}{2M_{34}^{2}}\left(M_{34}^{2}+m_{3}^{2}-m_{4}^{2}\right)\Big];\\
k_{4}^{\mu}&=\Big[\frac{\sqrt{M_{34}^{2}+X^{2}}}{2M_{34}^{2}}\left(M_{34}^{2}+m_{4}^{2}-m_{3}^{2}\right)+\frac{X}{2}\cos(\theta_{34})\lambda^{\frac{1}{2}}
\left(1,\frac{m_{3}^{2}}{M_{34}^{2}},\frac{m_{4}^{2}}{M_{34}^{2}}\right),\\
&-\frac{1}{2}M_{34}\lambda^{\frac{1}{2}}\left(1,\frac{m_{3}^{2}}{M_{34}^{2}},\frac{m_{4}^{2}}{M_{34}^{2}}\right)\sin(\theta_{34})\cos(\phi)
,-\frac{1}{2}M_{34}\lambda^{\frac{1}{2}}\left(1,\frac{m_{3}^{2}}{M_{34}^{2}},\frac{m_{4}^{2}}{M_{34}^{2}}\right)\sin(\theta_{34})\sin(\phi),\\
&-\frac{1}{2}\sqrt{M_{34}^{2}+X^{2}}\cos(\theta_{34})
\lambda^{\frac{1}{2}}\left(1,\frac{m_{3}^{2}}{M_{34}^{2}},\frac{m_{4}^{2}}{M_{34}^{2}}\right)-\frac{X}{2M_{34}^{2}}\left(M_{34}^{2}+m_{4}^{2}-m_{3}^{2}\right)\Big];
\end{split}
\end{equation}
where $X=\frac{1}{2m}\lambda^{\frac{1}{2}}\left(m^{2},M_{12}^{2},M_{34}^{2}\right)$.
The results for $B_{c}^{-}\rightarrow J/\psi\ell_{1}^{-}\ell_{2}^{-}\pi^{+}$ are obtained in an analogous way, although they are a bit more complicated due to the additional form factors involved in the pseudoscalar to vector meson transition.


\begin{thebibliography}{50}
\bibitem{neutrino oscillation 1} 
  Y.~Fukuda {\it et al.} [Super-Kamiokande Collaboration],
  Phys.\ Rev.\ Lett.\  {\bf 81}, 1562 (1998)
  [hep-ex/9807003].
  \bibitem{neutrino oscillation 2}
 S.~Fukuda {\it et al.} [Super-Kamiokande Collaboration],
  Phys.\ Rev.\ Lett.\  {\bf 86}, 5656 (2001)
  [hep-ex/0103033].
  \bibitem{neutrino oscillation 3} 
  S.~Fukuda {\it et al.} [Super-Kamiokande Collaboration],
  Phys.\ Lett.\ B {\bf 539}, 179 (2002)
  [hep-ex/0205075].
  \bibitem{neutrino oscillation 4} 
  Y.~Ashie {\it et al.} [Super-Kamiokande Collaboration],
  Phys.\ Rev.\ Lett.\  {\bf 93}, 101801 (2004)
  [hep-ex/0404034].
  \bibitem{neutrino oscillation 5} 
  B.~T.~Cleveland, T.~Daily, R.~Davis, Jr., J.~R.~Distel, K.~Lande, C.~K.~Lee, P.~S.~Wildenhain and J.~Ullman,
  Astrophys.\ J.\  {\bf 496}, 505 (1998).
  \bibitem{neutrino oscillation 6} 
  W.~Hampel {\it et al.} [GALLEX Collaboration],
  Phys.\ Lett.\ B {\bf 447}, 127 (1999).
  \bibitem{neutrino oscillation 7} 
  J.~N.~Abdurashitov {\it et al.} [SAGE Collaboration],
  J.\ Exp.\ Theor.\ Phys.\  {\bf 95}, 181 (2002)
  [Zh.\ Eksp.\ Teor.\ Fiz.\  {\bf 122}, 211 (2002)]
  [astro-ph/0204245].
  \bibitem{neutrino oscillation 8} 
  Q.~R.~Ahmad {\it et al.} [SNO Collaboration],
  Phys.\ Rev.\ Lett.\  {\bf 87}, 071301 (2001)
  [nucl-ex/0106015].
  \bibitem{neutrino oscillation 9} 
  Q.~R.~Ahmad {\it et al.} [SNO Collaboration],
  Phys.\ Rev.\ Lett.\  {\bf 89}, 011301 (2002)
  [nucl-ex/0204008].
  \bibitem{neutrino oscillation 10} 
  S.~N.~Ahmed {\it et al.} [SNO Collaboration],
  Phys.\ Rev.\ Lett.\  {\bf 92}, 181301 (2004)
  [nucl-ex/0309004].
  \bibitem{neutrino oscillation 11} 
  K.~Eguchi {\it et al.} [KamLAND Collaboration],
  Phys.\ Rev.\ Lett.\  {\bf 90}, 021802 (2003)
  [hep-ex/0212021].
  \bibitem{seesaw1} 
  P.~Minkowski,
  Phys.\ Lett.\ B {\bf 67}, 421 (1977).
\bibitem{seesaw2}T. Yanagida, in Proc. of the Workshop on Grand
Unified Theory and Baryon Number of the Universe, KEK, Japan, 1979.
\bibitem{seesaw3}M. Gell-Mann, P.Ramond and R. Slansky in Sanibel Symposium, February 1979, CALT-68-709 [retroprint
arXiv:hep-ph/9809459], and in Supergravity, eds. D. Freedman et al. (North Holland,
Amsterdam, 1979); S. L. Glashow in Quarks and Leptons, Cargese, eds. M. Levy et al.
(Plenum, 1980, New York), p. 707.
\bibitem{seesaw5} 
  R.~N.~Mohapatra and G.~Senjanovic,
  Phys.\ Rev.\ Lett.\  {\bf 44}, 912 (1980).
   \bibitem{Valle 1} 
   J.~Schechter and J.~W.~F.~Valle,
   Phys.\ Rev.\ D {\bf 22}, 2227 (1980).
   \bibitem{Valle 2} 
   J.~Schechter and J.~W.~F.~Valle,
   Phys.\ Rev.\ D {\bf 25}, 774 (1982).
  \bibitem{low energy seesaw1} 
  I.~Dorsner and P.~Fileviez Perez,
  JHEP {\bf 0706}, 029 (2007)
  [hep-ph/0612216];
   B.~Bajc, M.~Nemevsek and G.~Senjanovic,
  Phys.\ Rev.\ D {\bf 76}, 055011 (2007)
  [hep-ph/0703080].
  \bibitem{low energy seesaw2} 
  A.~de Gouvea, J.~Jenkins and N.~Vasudevan,
  Phys.\ Rev.\ D {\bf 75}, 013003 (2007)
  [hep-ph/0608147];
 A.~de Gouvea,
  arXiv:0706.1732 [hep-ph].
  \bibitem{lsnd} 
  C.~Athanassopoulos {\it et al.} [LSND Collaboration],
  Nucl.\ Instrum.\ Meth.\ A {\bf 388}, 149 (1997)
  [nucl-ex/9605002].
  \bibitem{lsnd1} 
  J.~M.~Conrad, W.~C.~Louis and M.~H.~Shaevitz,
  Ann.\ Rev.\ Nucl.\ Part.\ Sci.\  {\bf 63}, 45 (2013)
  [arXiv:1306.6494 [hep-ex]].
  \bibitem{miniboone} 
  A.~A.~Aguilar-Arevalo {\it et al.} [MiniBooNE Collaboration],
  Phys.\ Rev.\ Lett.\  {\bf 98}, 231801 (2007)
  [arXiv:0704.1500 [hep-ex]].
  \bibitem{miniboone1} 
  A.~A.~Aguilar-Arevalo {\it et al.} [MiniBooNE Collaboration],
  Phys.\ Rev.\ Lett.\  {\bf 103}, 111801 (2009)
  [arXiv:0904.1958 [hep-ex]].
  \bibitem{miniboone2} 
  A.~A.~Aguilar-Arevalo {\it et al.} [MiniBooNE Collaboration],
  Phys.\ Rev.\ Lett.\  {\bf 110}, 161801 (2013)
  [arXiv:1207.4809 [hep-ex], arXiv:1303.2588 [hep-ex]].
  \bibitem{chooz} 
  Y.~Abe {\it et al.} [Double Chooz Collaboration],
  Phys.\ Rev.\ D {\bf 86}, 052008 (2012)
  [arXiv:1207.6632 [hep-ex]].
  \bibitem{dayabay} 
  F.~P.~An {\it et al.} [Daya Bay Collaboration],
  Phys.\ Rev.\ Lett.\  {\bf 108}, 171803 (2012)
  [arXiv:1203.1669 [hep-ex]].
  \bibitem{reno} 
  J.~K.~Ahn {\it et al.} [RENO Collaboration],
  Phys.\ Rev.\ Lett.\  {\bf 108}, 191802 (2012)
  [arXiv:1204.0626 [hep-ex]].
  \bibitem{kev sterile1} 
  S.~Dodelson and L.~M.~Widrow,
  Phys.\ Rev.\ Lett.\  {\bf 72}, 17 (1994)
  [hep-ph/9303287].
  \bibitem{kev sterile2} 
  X.~D.~Shi and G.~M.~Fuller,
  Phys.\ Rev.\ Lett.\  {\bf 82}, 2832 (1999)
  [astro-ph/9810076];
   K.~Abazajian, G.~M.~Fuller and M.~Patel,
  Phys.\ Rev.\ D {\bf 64}, 023501 (2001)
  [astro-ph/0101524];
  K.~N.~Abazajian and G.~M.~Fuller,
  Phys.\ Rev.\ D {\bf 66}, 023526 (2002)
  [astro-ph/0204293].
  G.~M.~Fuller, A.~Kusenko, I.~Mocioiu and S.~Pascoli,
  Phys.\ Rev.\ D {\bf 68}, 103002 (2003)
  [astro-ph/0307267].
   K.~Abazajian,
  Phys.\ Rev.\ D {\bf 73}, 063506 (2006)
  [astro-ph/0511630].
  \bibitem{kev sterile3} 
  A.~Kusenko,
  Phys.\ Rept.\  {\bf 481}, 1 (2009)
  [arXiv:0906.2968 [hep-ph]];
   A.~Kusenko,
  Int.\ J.\ Mod.\ Phys.\ D {\bf 16}, 2325 (2008)
  [astro-ph/0608096];
  T.~Asaka, M.~Shaposhnikov and A.~Kusenko,
  Phys.\ Lett.\ B {\bf 638}, 401 (2006)
  [hep-ph/0602150].
  P.~L.~Biermann and A.~Kusenko,
  Phys.\ Rev.\ Lett.\  {\bf 96}, 091301 (2006)
  [astro-ph/0601004].
  \bibitem{kev sterile4} 
  A.~Boyarsky, O.~Ruchayskiy and M.~Shaposhnikov,
  Ann.\ Rev.\ Nucl.\ Part.\ Sci.\  {\bf 59}, 191 (2009)
  [arXiv:0901.0011 [hep-ph]];
M.~Shaposhnikov and I.~Tkachev,
  Phys.\ Lett.\ B {\bf 639}, 414 (2006)
  [hep-ph/0604236].
  \bibitem{kev sterile5}
   H.~J.~de Vega and N.~G.~Sanchez,
  Mon.\ Not.\ Roy.\ Astron.\ Soc.\  {\bf 404}, 885 (2010)
  [arXiv:0901.0922 [astro-ph.CO]];
  H.~J.~de Vega, P.~Salucci and N.~G.~Sanchez,
  New Astron.\  {\bf 17}, 653 (2012)
  [arXiv:1004.1908 [astro-ph.CO]];
  H.~J.~de Vega and N.~G.~Sanchez,
  arXiv:1109.3187 [astro-ph.CO];
  C.~Destri, H.~J.~de Vega and N.~G.~Sanchez,
  New Astron.\  {\bf 22}, 39 (2013)
  [arXiv:1204.3090 [astro-ph.CO]].
\bibitem{kevsterileWP} R.~Adhikari {\it et al.},
arXiv:1602.04816 [hep-ph].
\bibitem{Queiroz} 
F.~S.~Queiroz and K.~Sinha,
Phys.\ Lett.\ B {\bf 735}, 69 (2014)
doi:10.1016/j.physletb.2014.06.016
[arXiv:1404.1400 [hep-ph]].
\bibitem{peak searches} 
  R.~E.~Shrock,
  Phys.\ Lett.\ B {\bf 96}, 159 (1980).
  \bibitem{tao han} 
  A.~Atre, T.~Han, S.~Pascoli and B.~Zhang,
  JHEP {\bf 0905}, 030 (2009)
  [arXiv:0901.3589 [hep-ph]].
  \bibitem{kim2} 
  G.~Cvetic, C.~Dib, S.~K.~Kang and C.~S.~Kim,
  Phys.\ Rev.\ D {\bf 82}, 053010 (2010)
  [arXiv:1005.4282 [hep-ph]]; G.~Cvetic, C.~Dib and C.~S.~Kim,
  JHEP {\bf 1206}, 149 (2012)
  [arXiv:1203.0573 [hep-ph]].
  \bibitem{GammaN calculation} 
  J.~C.~Helo, S.~Kovalenko and I.~Schmidt,
  Nucl.\ Phys.\ B {\bf 853}, 80 (2011)
  [arXiv:1005.1607 [hep-ph]].
  \bibitem{cleo collaboration1} 
  Q.~He {\it et al.} [CLEO Collaboration],
  Phys.\ Rev.\ Lett.\  {\bf 95}, 221802 (2005)
  [hep-ex/0508031].
  \bibitem{cleo collaboration2} 
  Y.~Kubota {\it et al.} [CLEO Collaboration],
  Nucl.\ Instrum.\ Meth.\ A {\bf 320}, 66 (1992).
  \bibitem{cleo collaboration3} 
  P.~Rubin {\it et al.} [CLEO Collaboration],
  Phys.\ Rev.\ D {\bf 82}, 092007 (2010)
  [arXiv:1009.1606 [hep-ex]].
  \bibitem{focus} 
  J.~M.~Link {\it et al.} [FOCUS Collaboration],
  Phys.\ Lett.\ B {\bf 572}, 21 (2003)
  [hep-ex/0306049].
  \bibitem{babar} 
  J.~P.~Lees {\it et al.} [BaBar Collaboration],
  Phys.\ Rev.\ D {\bf 84}, 072006 (2011)
  [arXiv:1107.4465 [hep-ex]].
  \bibitem{lhcb1} 
  R.~Aaij {\it et al.} [LHCb Collaboration],
  Phys.\ Rev.\ Lett.\  {\bf 108}, 101601 (2012)
  [arXiv:1110.0730 [hep-ex]];
M. Patel, talk given at the
Workshop on Flavor and the Fourth Family, IPPP Durham 14-16 September (2011).
\bibitem{Dittmar:1989yg} 
M.~Dittmar, A.~Santamaria, M.~C.~Gonzalez-Garcia and J.~W.~F.~Valle,
Nucl.\ Phys.\ B {\bf 332}, 1 (1990).
\bibitem{delAguila:2006bda} 
F.~del Aguila, J.~A.~Aguilar-Saavedra and R.~Pittau,
J.\ Phys.\ Conf.\ Ser.\  {\bf 53}, 506 (2006), [hep-ph/0606198].
\bibitem{Buchmuller:1991tu} 
W.~Buchmuller and C.~Greub,
Nucl.\ Phys.\ B {\bf 363}, 345 (1991).
\bibitem{Almeida:2000yx} 
F.~M.~L.~Almeida, Jr., Y.~D.~A.~Coutinho, J.~A.~Martins Simoes and M.~A.~B.~do Vale,
Phys.\ Rev.\ D {\bf 63}, 075005 (2001).
\bibitem{Neutrino and collider physics} 
 F.~F.~Deppisch, P.~S.~Bhupal Dev and A.~Pilaftsis,
  New J.\ Phys.\  {\bf 17}, no. 7, 075019 (2015)
  [arXiv:1502.06541 [hep-ph]].
\bibitem{Prospects of Heavy Neutrino Searches at Future Lepton Colliders} 
 S.~Banerjee, P.~S.~B.~Dev, A.~Ibarra, T.~Mandal and M.~Mitra,
  Phys.\ Rev.\ D {\bf 92}, 075002 (2015)
  [arXiv:1503.05491 [hep-ph]].
  \bibitem{New Production Mechanism for Heavy Neutrinos at the LHC} 
  P.~S.~B.~Dev, A.~Pilaftsis and U.~k.~Yang,
  Phys.\ Rev.\ Lett.\  {\bf 112}, no. 8, 081801 (2014)
  [arXiv:1308.2209 [hep-ph]].
    \bibitem{Direct bounds on electroweak scale pseudo-Dirac neutrinos from 8 TeV LHC data} 
  A.~Das, P.~S.~Bhupal Dev and N.~Okada,
  Phys.\ Lett.\ B {\bf 735}, 364 (2014)
  [arXiv:1405.0177 [hep-ph]].
\bibitem{Bray:2005wv} 
S.~Bray, J.~S.~Lee and A.~Pilaftsis,
Phys.\ Lett.\ B {\bf 628}, 250 (2005).
\bibitem{ppcollider2}W.-Y. Keung and G. Senjanovic, Phys. Rev. Lett. 50, 1427 (1983); 
T.~H.~Ho, C.~R.~Ching and Z.~J.~Tao,
Phys.\ Rev.\ D {\bf 42}, 2265 (1990);D.~A.~Dicus, D.~D.~Karatas and P.~Roy,
Phys.\ Rev.\ D {\bf 44}, 2033 (1991);
A.~Datta, M.~Guchait and D.~P.~Roy,
Phys.\ Rev.\ D {\bf 47}, 961 (1993);
A.~Ferrari, J.~Collot, M.~L.~Andrieux, B.~Belhorma, P.~de Saintignon, J.~Y.~Hostachy, P.~Martin and M.~Wielers,
Phys.\ Rev.\ D {\bf 62}, 013001 (2000).
\bibitem{ppcollider1} 
  A.~Ali, A.~V.~Borisov and N.~B.~Zamorin,
  Eur.\ Phys.\ J.\ C {\bf 21}, 123 (2001)
  [hep-ph/0104123].
\bibitem{Almeida:2000pz} 
F.~M.~L.~Almeida, Jr., Y.~D.~A.~Coutinho, J.~A.~Martins Simoes and M.~A.~B.~do Vale,
Phys.\ Rev.\ D {\bf 62}, 075004 (2000).

\bibitem{ppbarcollider1} 
  T.~Han and B.~Zhang,
  Phys.\ Rev.\ Lett.\  {\bf 97}, 171804 (2006)
  [hep-ph/0604064].
\bibitem{Rizzo:1982kn} 
T.~G.~Rizzo,
Phys.\ Lett.\ B {\bf 116}, 23 (1982); C.~A.~Heusch and P.~Minkowski,
Nucl.\ Phys.\ B {\bf 416}, 3 (1994).
\bibitem{deshpande} 
S.~Bar-Shalom, N.~G.~Deshpande, G.~Eilam, J.~Jiang and A.~Soni,
Phys.\ Lett.\ B {\bf 643}, 342 (2006).
 \bibitem{fourbody decay} 
 N.~Quintero, G.~Lopez Castro and D.~Delepine,
 Phys.\ Rev.\ D {\bf 84}, 096011 (2011)
 [Phys.\ Rev.\ D {\bf 86}, 079905 (2012)]
 [arXiv:1108.6009 [hep-ph]].
\bibitem{lhcb} 
 R.~Aaij {\it et al.} [LHCb Collaboration],
 Phys.\ Rev.\ Lett.\  {\bf 111}, no. 18, 181801 (2013).
 \bibitem{atlas} 
  G.~Aad {\it et al.} [ATLAS Collaboration],
  Eur.\ Phys.\ J.\ C {\bf 76}, no. 1, 4 (2016)
  [arXiv:1507.07099 [hep-ex]].
  \bibitem{cms} 
  V.~Khachatryan {\it et al.} [CMS Collaboration],
  JHEP {\bf 1501}, 063 (2015)
  [arXiv:1410.5729 [hep-ex]].
\bibitem{BcProd}
R.~Aaij {\it et al.} [LHCb Collaboration],
Phys.\ Rev.\ Lett.\  {\bf 109}, 232001 (2012);  N.~Brambilla {\it et al.} [Quarkonium Working Group Collaboration],
hep-ph/0412158.

  \bibitem{form factors} 
 V.~V.~Kiselev, A.~E.~Kovalsky and A.~K.~Likhoded,
  Nucl.\ Phys.\ B {\bf 585}, 353 (2000)
  [hep-ph/0002127], V. V. Kiselev, hep-ph/0211021.
  \bibitem{PDG} 
  K.~A.~Olive {\it et al.} [Particle Data Group Collaboration],
  Chin.\ Phys.\ C {\bf 38}, 090001 (2014).
  \bibitem{Global Analyses of Neutrino Oscillation Experiments} 
  M.~C.~Gonzalez-Garcia, M.~Maltoni and T.~Schwetz,
  arXiv:1512.06856 [hep-ph].
  \bibitem{Unitarity and the three flavour neutrino mixing matrix} 
  S.~Parke and M.~Ross-Lonergan,
  arXiv:1508.05095 [hep-ph].
  \bibitem{four-body tau} 
  G.~Lopez Castro and N.~Quintero,
  Phys.\ Rev.\ D {\bf 85}, 076006 (2012)
  [Phys.\ Rev.\ D {\bf 86}, 079904 (2012)]
  [arXiv:1203.0537 [hep-ph]].
   \bibitem{Yuan}
  H.~Yuan, T.~Wang, G.~L.~Wang, W.~L.~Ju and J.~M.~Zhang,
  JHEP {\bf 1308}, 066 (2013)
  [arXiv:1304.3810 [hep-ph]].
  \bibitem{Dong}
  H.~R.~Dong, F.~Feng and H.~B.~Li,
  Chin.\ Phys.\ C {\bf 39}, no. 1, 013101 (2015)
  [arXiv:1305.3820].
  \bibitem{Lepton number violating decay} 
 G.~Lopez Castro and N.~Quintero,
  Nucl.\ Phys.\ Proc.\ Suppl.\  {\bf 253-255}, 12 (2014)
  [arXiv:1212.0037 [hep-ph]].
   \bibitem{LHCb2012}
   R.~Aaij {\it et al.} [LHCb Collaboration],
   Phys.\ Rev.\ D {\bf 85}, 112004 (2012)
   [arXiv:1201.5600 [hep-ex]].
   \bibitem{Chang} 
  C.~H.~Chang, C.~Driouichi, P.~Eerola and X.~G.~Wu,
  Comput.\ Phys.\ Commun.\  {\bf 159}, 192 (2004)
  [hep-ph/0309120].
   \bibitem{Belyaev} Our crude estimate is based on private communication with Vanya Belyaev from the LHCb collaboration.
   \bibitem{BcProdLHCb}
   R.~Aaij {\it et al.} [LHCb Collaboration],
   Phys.\ Rev.\ Lett.\  {\bf 114}, 132001 (2015)
   [arXiv:1411.2943 [hep-ex]].
  
  
 
  
  
  \bibitem{Global Constraints on a Heavy Neutrino} 
  A.~de Gouvêa and A.~Kobach,
  Phys.\ Rev.\ D {\bf 93}, no. 3, 033005 (2016)
  [arXiv:1511.00683 [hep-ph]].
 
  \bibitem{CHARM1} 
  P.~Vilain {\it et al.} [CHARM II Collaboration],
  Phys.\ Lett.\ B {\bf 343}, 453 (1995)
  [Phys.\ Lett.\ B {\bf 351}, 387 (1995)].
\bibitem{CHARM2} 
  J.~Orloff, A.~N.~Rozanov and C.~Santoni,
  Phys.\ Lett.\ B {\bf 550}, 8 (2002)
  [hep-ph/0208075].
  \bibitem{NOMAD} 
  P.~Astier {\it et al.} [NOMAD Collaboration],
  Phys.\ Lett.\ B {\bf 506}, 27 (2001)
  [hep-ex/0101041].
  \bibitem{DELPHI} 
  P.~Abreu {\it et al.} [DELPHI Collaboration],
  Z.\ Phys.\ C {\bf 74}, 57 (1997)
  Erratum: [Z.\ Phys.\ C {\bf 75}, 580 (1997)].
 
\end{thebibliography}
\end{document}